# Identification of images of COVID-19 from Chest Computed Tomography (CT) scans using Deep learning: Comparing COGNEX VisionPro Deep Learning 1.0™ Software with Open Source Convolutional Neural Networks

Arjun Sarkar[1,2,*], Joerg Vandenhirtz, PhD[2], Jozsef Nagy[2], David Bacsa[2], Mitchell Riley[2]

1. FH Aachen University of Applied Sciences, Department of Biomedical Engineering, Germany,
2. COGNEX Corporation, which has funded the research in this paper.

---

**ABSTRACT**

The COVID-19 pandemic has been having a devastating effect on human life. For screening and testing patients, along with RT-PCR (Real-time Reverse Transcriptase-Polymerase Chain Reaction) testing, chest radiology images are being used. For detection of COVID-19 from radiology images, many organizations are proposing the use of Deep Learning for automating the process. University of Waterloo and Darwin AI – a startup spinoff from the University, have designed their own Deep Learning model 'COVIDNet-CT' to detect COVID-19 from infected chest CT images. Additionally, they have introduced a CT image dataset 'COVIDx-CT', from CT images collected by the China national Center for Bioinformation. COVIDx-CT contains 104,009 CT image slices across 1,489 patient cases. After obtaining remarkable results on identifying of images of COVID-19 from chest X-ray images by using the COGNEX VisionPro Deep Learning Software 1.0 [1] this time we test the performance of the software on identification of images of COVID-19 from CT scans. COGNEX's Deep Learning Software: VisionPro Deep Learning™, is a Deep Learning software which is used across various domains ranging from factory automation to life sciences. In this study we train the classification model on 82,818 chest CT training and validation images from the COVIDx-CT dataset in 3 classes: normal, pneumonia and COVID-19 and then test the results of the classification on the 21,191 test images are compared with the results of COVIDNet-CT and various other state of the art Deep Learning models from the open-source community. Also, we test how reducing the number of images in the training set effects the results of the software. Overall, VisionPro Deep Learning gives the best results with F-scores over 99%, even as the number of images in the training set are reduced significantly.

**Abbreviations:**

CT – Computed Tomography
COVID-19 – Coronavirus Disease 2019
RNA – Ribonucleic Acid
ReLU – Rectified Linear Unit
RRT-PCR– Real-time Reverse Transcriptase-Polymerase Chain Reaction
GUI – Graphical User Interface
PPV – Positive Predictive Value

**INTRODUCTION**

The novel coronavirus disease – named COVID-19 by the World Health Organization– is caused by a new coronavirus class known as the SARS-CoV2 (Severe Acute Respiratory Syndrome Coronavirus 2). It is a single-stranded RNA (Ribonucleic Acid) virus that causes severe respiratory infections. The first COVID-19 cases were reported in December 2019, in Wuhan, Hubei province, China [1]. Since then, the virus has spread worldwide, and it has been given the status of a pandemic by the World Health Organization. As of 15th September 2020, 09:30 GMT, 29,4 million people have been infected, and 933 thousand people have died due to COVID-19 [2]. There have been no vaccines available, so far, for treating COVID-19. One of the best solutions has been detecting the virus in its early stages and then isolating the infected people by quarantining them, thus preventing healthy people from getting infected.

In many cases, Real-time Reverse Transcriptase-Polymerase Chain Reaction (RRT-PCR) of nasopharyngeal swabs have been used for diagnosis [3]. The RT-PCR throat swabs are collected from patients with COVID-19, and the RNA is then

---

[1] Arjun Sarkar, Joerg Vandenhirtz, Jozsef Nagy, David Bacsa and Mitchell Riley, Identification of images of COVID-19 from Chest X-rays using Deep Learning: Comparing COGNEX VisionPro Deep Learning 1.0 Software with Open Source Convolutional Neural Networks, arXiv:2008.00597v2 [eess. IV], 2020

extracted. This process takes over two hours to complete and has a long turnaround time with limited sensitivity. The best alternative is to detect images of COVID-19 from radiology scans [4,5,6] (chest X-Ray images and chest Computed Tomography (CT) images).

Radiologic examinations, especially thin-slice chest CT scans, play an important role in fighting this infectious disease [7]. On a study on 1014 patients from Wuhan, China, 55% patient had positive RT-PCR results, and 88% had positive Chest CT scan results [8]. The sensitivity of Chest CT scans in suggesting positive COVID-19 was 97%, based on positive RT-PCR results. Out of the 413 patients with negative RT-PCR results, 75% had positive CT findings. Of those 308 patients, 48% were considered highly likely cases and 33% to be probable cases. Thus, suggests that Chest CT scans may be considered as a primary tool for the current COVID-19 detection in epidemic areas. Chest CT scans can also help in identifying early phase lung infection [9,10]. Thin slice chest CT images are proven to be effective in early detection of COVID-19 [11,12].

About 20% of the patients infected with COVID-19 develop pulmonary infiltrates and some develop very serious abnormalities [13]. The virus reaches the lungs' gas exchange units and infects alveolar type 2 cells [14,15]. The most frequent CT abnormalities observed are ground-glass opacity, consolidation, and interlobular septal thickening in both lungs [16]. To detect these abnormalities, a significant number of expert radiologists who can interpret these radiology images are needed. Due to the ever-increasing number of cases of COVID-19 infections, it is getting harder for radiologists to keep up with this demand. In this scenario, Deep Learning techniques prove to be beneficial in both classifying the abnormalities from lung X-ray images and in aiding the radiologists to accurately predict COVID-19 cases in a reduced time frame.

While many studies have demonstrated success in detecting images of COVID-19 using Deep Learning with both CT scans and X-rays, most of the Deep Learning architectures need extensive programming. Due to the absence of a GUI (Graphical User Interface) with most of these Deep Learning models, it is difficult for radiologists, who lack knowledge in Deep Learning or programming, to use these models, let alone train them. Therefore, we showcase an already existing Deep Learning software with a very intuitive GUI, which can be used as a pretrained software or can even be trained on new data from hospitals or research centers.

COGNEX VisionPro Deep Learning™ is a Deep Learning vision software, from COGNEX Corporation (Headquarters: Natick, MA, United States). It is a field-tested, optimized, and reliable software solution based on a state-of-the-art set of machine learning and deep learning algorithms. VisionPro Deep Learning combines a comprehensive machine vision tool library with advanced deep learning tools.

In a previous study, we used VisionPro Deep Learning software to identify images of COVID-19 from Chest X-rays and got excellent results [17]. To further the study, we now test the software on Chest CT images. The VisionPro Deep Learning GUI "COGNEX Deep Learning Studio" has three tools for image classification, segmentation, and location. It contains various Deep Learning architectures built within the GUI, to carry out specific tasks:

1) Green Tool – This is the Classification tool. It is used to classify objects or complete scenes. It can be used to classify defects, cell types, images of different labels, or different types of test tubes used in laboratories. The Green tool learns from the collection of labeled images of different classes and can then be used to classify images that it has not seen previously. This tool is comparable to classification neural networks such as VGG [18], ResNet [19] or DenseNet [20].

2) Red Tool – This is the Analyze tool. It is used for segmentation and defect/anomaly detection, for example, to detect anomalies in blood samples (clots), incomplete or improper centrifugation, or sample quality management. The Red tool is also used to segment specific regions, such as defects or areas of interest. The Red tool comes with the option of using either Supervised Learning or Unsupervised Learning for segmentation and detection. This is comparable to the segmentation neural network, such as U-Net [21].

3) Blue Tool –

    a) This is the Feature Localization and Identification tool. The Blue tool finds complex features and objects by learning from labeled images. It has self-learning algorithms that can locate, classify, and count the objects in an image. It can be used for locating organs in X-ray images or cells on a microscopic slide.

b) The Blue tool also has Read feature. It is a pretrained model that helps decipher severely deformed and poorly etched words and codes using optical character recognition (OCR). This is the only pretrained tool. All the other tools need to be trained on images first to get results.

For the classification of Chest CT images, we use the Green (classification) tool of the VisionPro Deep Learning software.

## METHOD

### 1. Dataset

The original dataset of all the CT images and their metadata is obtained from the dataset created by the China Consortium of Chest CT image Investigation (CC-CCII) [22] and is provided by the China National Center for Bioinformation(CNCB) and can be found at the following link - http://ncov-ai.big.ac.cn. In this dataset, CT images are classified into three classes – novel coronavirus pneumonia (NCP) caused by the SARS-CoV-2 virus, common pneumonia (CP) and normal controls. The dataset is publicly available, and it is aimed at aiding clinicians and researchers to combat the COVID-19 pandemic.

From this dataset researchers at the University of Waterloo and Darwin AI came up with the COVIDx-CT dataset [23]. This dataset is publicly available under the following link - https://github.com/haydengunraj/COVIDNet-CT [29]. The Waterloo team trained their own deep learning neural network, called COVIDNet-CT on this dataset. The dataset contains 104,009 CT slices from 1,489 patients. The dataset is divided into a train, validation and test set. Each of these sets contain images belonging to three classes, Normal (CT slices of patients without any kind of infection), Pneumonia (CT slices of patients infected by various bacterial/viral pneumonia but not with COVID-19) and COVID-19 (CT slices of patients infected with COVID-19).

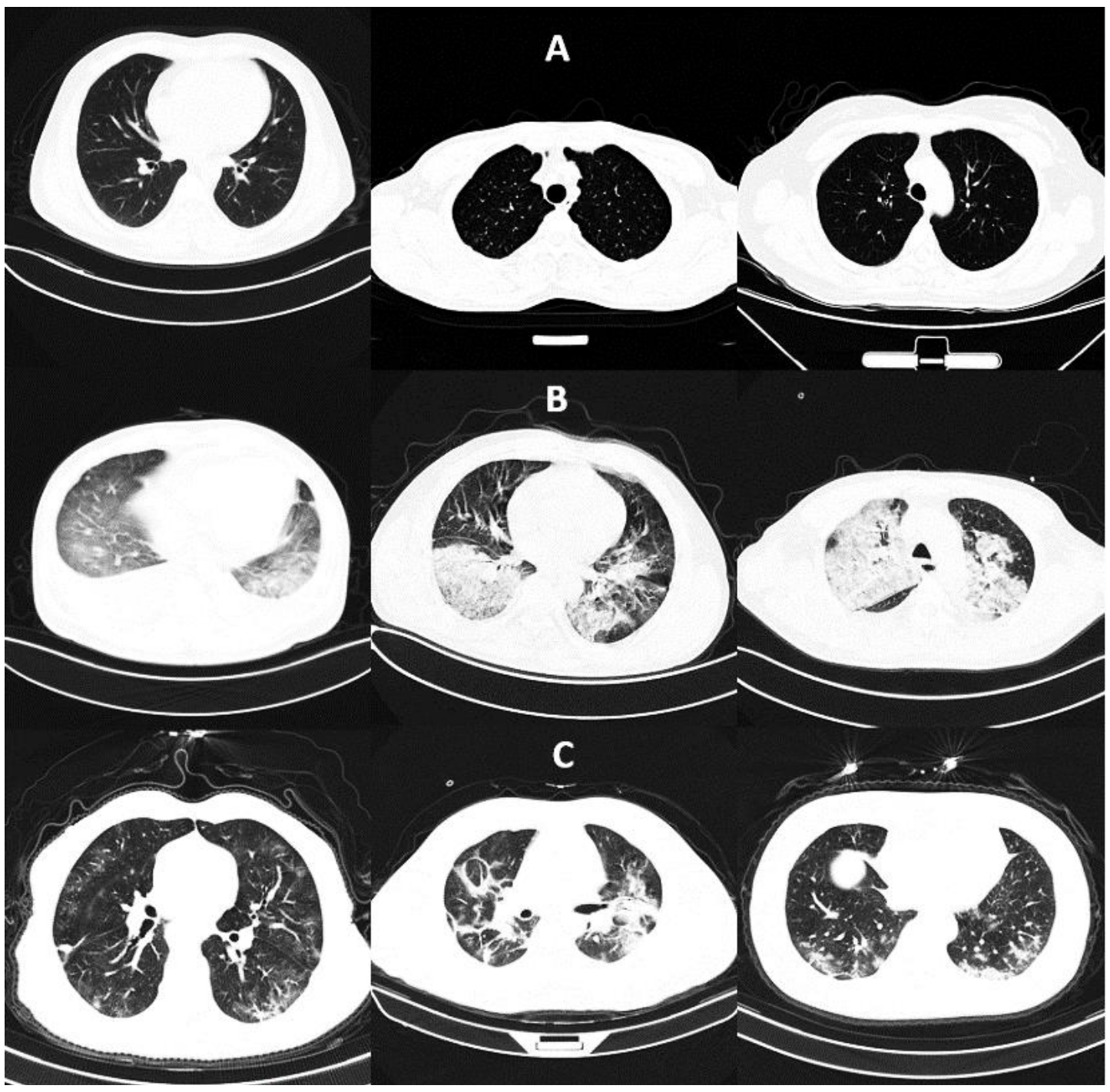

**Figure 1:** Examples of the chest CT images belonging to the different classes. The class numbers are shown along the vertical axis. Class A: Normal images, Class B: Pneumonia images and Class 3: COVID-19 images. All the images belong to the training set of the COVIDx-CT dataset [18].

We further classify the COVIDx-CT [23] dataset into three settings. In each setting, we decrease the number of training images and increase the number of test images and check the performance of VisionPro Deep Learning and various open source models, as the number of training images decreases.

**Setting 1:**

In Setting 1, the COVIDx-CT [23] dataset is used as it is, to enable a 1 to 1 comparison with the COVID-Net results It contains a total of 61,782 CT slices in the training set, 21,036 CT slices in the validation set and 21,191 CT slices in the test set. The number of images in each class in each of these sets is shown in Table 1.

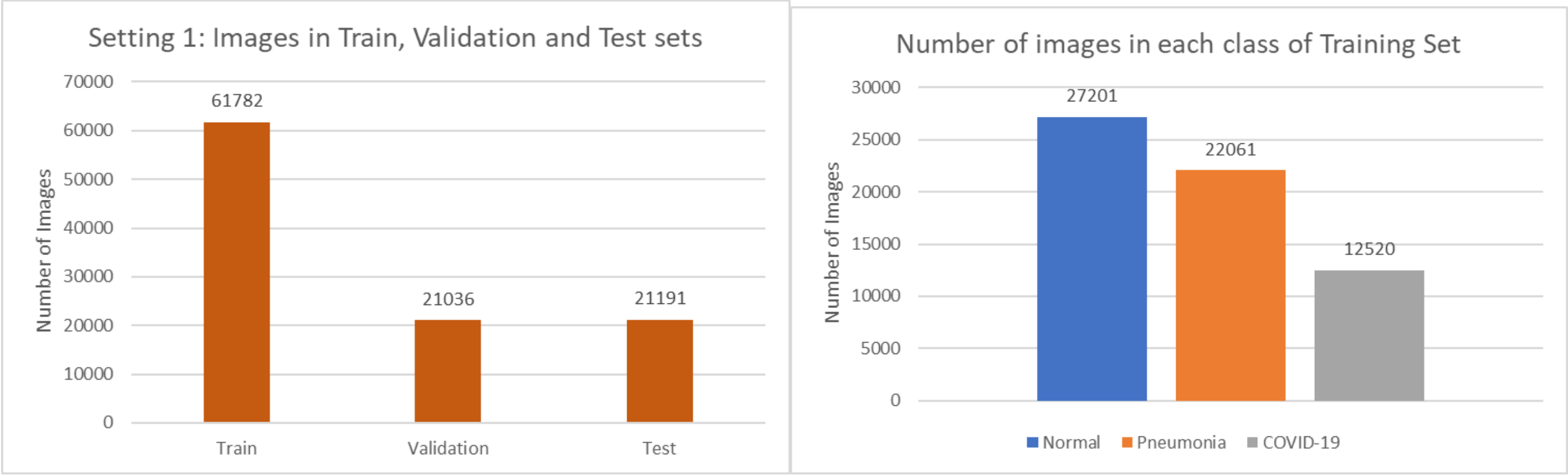

**Figure 2:** Left: Chest CT image distribution across the train, test and validation sets for Setting 1. Right: Distribution of images in classes Normal, Pneumonia and COVID-19 in the training set of Setting 1

| | Normal | Pneumonia | COVID-19 | Total |
|---|---|---|---|---|
| **Train** | 27201 | 22061 | 12520 | **61782** |
| **Validation** | 9107 | 7400 | 4529 | **21036** |
| **Test** | 9450 | 7395 | 4346 | **21191** |

**Table 1:** Number of images belonging to each class for Setting 1

**Setting 2:**

In Setting 2, approximately nine thousand images are moved from the training set to the test set. The validation set is left as it is. This is the first setting in which the number of images is reduced in the training set and thus lower than the original COVIDx-CT [23] dataset. The idea behind this is to check how reducing the number of images in the training set (and at the same time increasing the number of test images) affects the final results. While transferring images from the training set to the test set, it is made sure that patient information is preserved, and there are no overlapping CT slices both in the test and the train set. This is done to prevent any kind of overfitting. Also, approximately equal number of images are transferred from each class. The number of images in each class of each set is shown in Table 2.

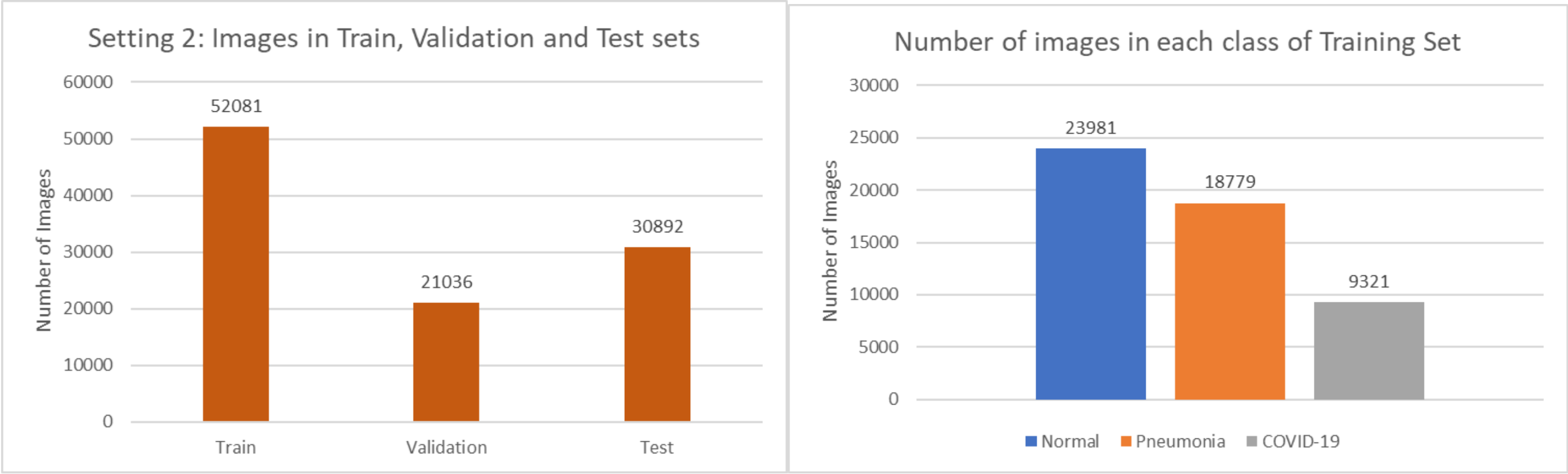

**Figure 3:** Left: Chest CT image distribution across the train, test and validation sets for Setting 2. Right: Distribution of images in classes Normal, Pneumonia and COVID-19 in the training set of Setting 2

| | Normal | Pneumonia | COVID-19 | Total |
|---|---|---|---|---|
| **Train** | 23981 | 18779 | 9321 | **52081** |
| **Validation** | 9107 | 7400 | 4529 | **21036** |
| **Test** | 12670 | 10677 | 7545 | **30892** |

**Table 2:** Number of images belonging to each class for Setting 2

**Setting 3:**

In Setting 3, even more images are moved from the training set to the test set, taking the total number of images in the test set to 56,635, as compared to only 21,191 test images in the original COVIDx-CT [23] images. The validation set is left undisturbed. As in Setting 2, even with Setting 3, it is made sure that one patients CT slices do not fall both in the train and test set. The number of images in each class of each set is shown in Table 3.

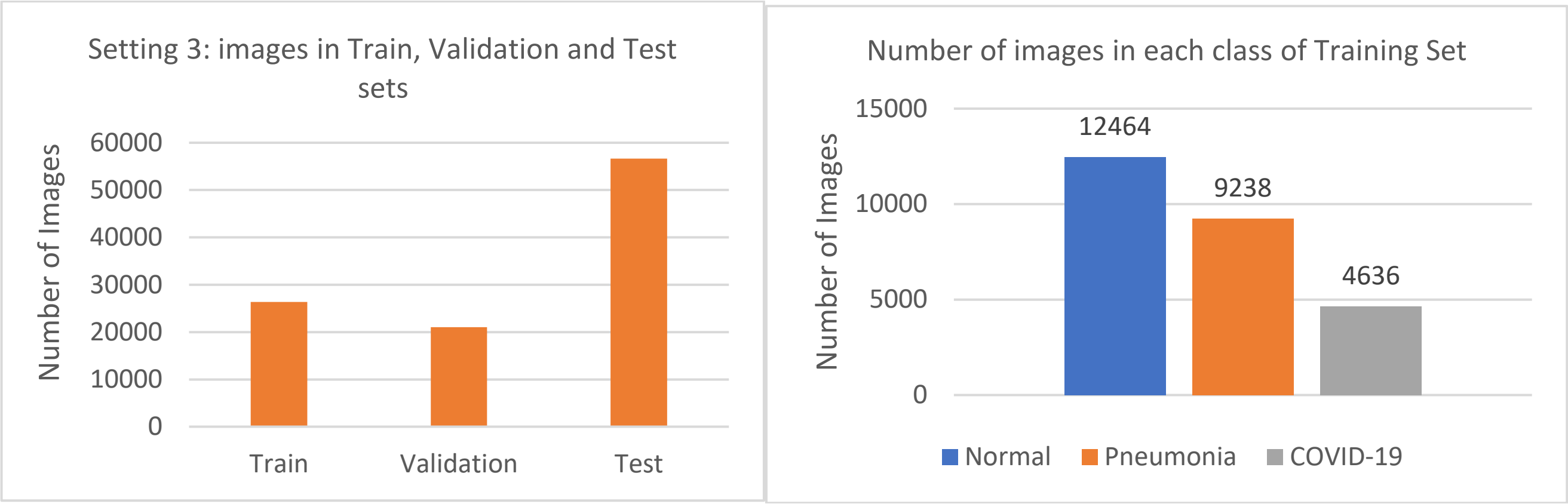

**Figure 4:** Left: Chest CT image distribution across the train, test and validation sets for Setting 3. Right: Distribution of images in classes Normal, Pneumonia and COVID-19 in the training set of Setting 3

| | Normal | Pneumonia | COVID-19 | Total |
|---|---|---|---|---|
| **Train** | 12464 | 9238 | 4636 | **26338** |
| **Validation** | 9107 | 7400 | 4529 | **21036** |
| **Test** | 24187 | 20218 | 12230 | **56635** |

**Table 3:** Number of images belonging to each class for Setting 3

**2. Preprocessing**

The scripts for generating the COVIDx-CT dataset provided at https://github.com/haydengunraj/COVIDNet-CT [29] and are used to create the original COVIDx-CT dataset [23]. The dataset separates the images into train, validation and test folders. Along with the images, the script also generates three text files containing the names of images belonging to the train, validation and test folders, and their class labels.

To simplify classification, a python script is used to convert the '.txt' files into 'pandas' data frames and then finally converted to '.csv' files for better understanding. Next, another python script is created to rename all the CT images of the train, validation and test folders according to their class labels and store them in a new train, validation and test directories. Since the goal is classification of the CT images, renaming the images makes it easier to interpret the images directly from their file names, rather than consulting a '.csv' file every time.

Unlike most other Deep Learning architectures VisionPro Deep Learning does not need any preprocessing of the images. The images can be fed directly into the GUI, and the software automatically does the preprocessing, before starting to train the model. No other preprocessing steps are necessary with VisionPro Deep learning, such as - setting class weights or oversampling of the imbalanced classes, which are necessary for training the other open-source CNN models. Once the images are fed into the VisionPro Deep learning GUI, they are ready to be trained.

For all the other open-source Deep Learning architectures used, it is necessary to first preprocess the images before feeding it into the neural network architecture. The images are first resized to a size of 224 x 224 pixels. The images are also standardized and mapped between [-1,1]. This is done, keeping in mind that standardization helps the Deep Learning network to learn much faster. The entire training is done on a Nvidia 2080 GPU, and as to not run into 'GPU memory errors' this is found to be the perfect image size. To address the class imbalance problem, class weights are generated using the 'scikit-learn' [24] library. These class weights are then applied while fitting the model for training the neural network.

Each of the Deep Learning architectures are run three times for each of the three settings as mentioned in Dataset part (part 1) of the Method section.

**3. Classification using VisionPro Deep Learning**

The goal of the study is the classification of Normal, Pneumonia and COVID-19 CT images. For classification, VisionPro Deep Learning uses the Green tool. Once the images are loaded and labeled, they are ready for training. In VisionPro Deep Learning, the Region of Interest (ROI) of the images can be selected. Thus, it is possible to reduce the edges by 10-20% to remove artifacts like the letters or borders, which are usually at the edges of the images. In this case, the entire images are used without cropping the edges because many images have the lungs towards the edges, and we did not want to remove any essential information.

For feeding the images into VisionPro Deep learning, the images do not need to be resized. Images of all resolutions and aspect ratio can be fed into the GUI, and the GUI does the preprocessing automatically before starting the training. In VisionPro Deep learning, the Green tool has two subcategories, 'High-detail' and 'Focused'. Under High-detail there are several options such as sizes of model architectures - Small, Normal, Large and Extra-Large models, which can be selected for training the model. We train the network using the 'High detail' subcategory and selecting the 'Normal' size model.

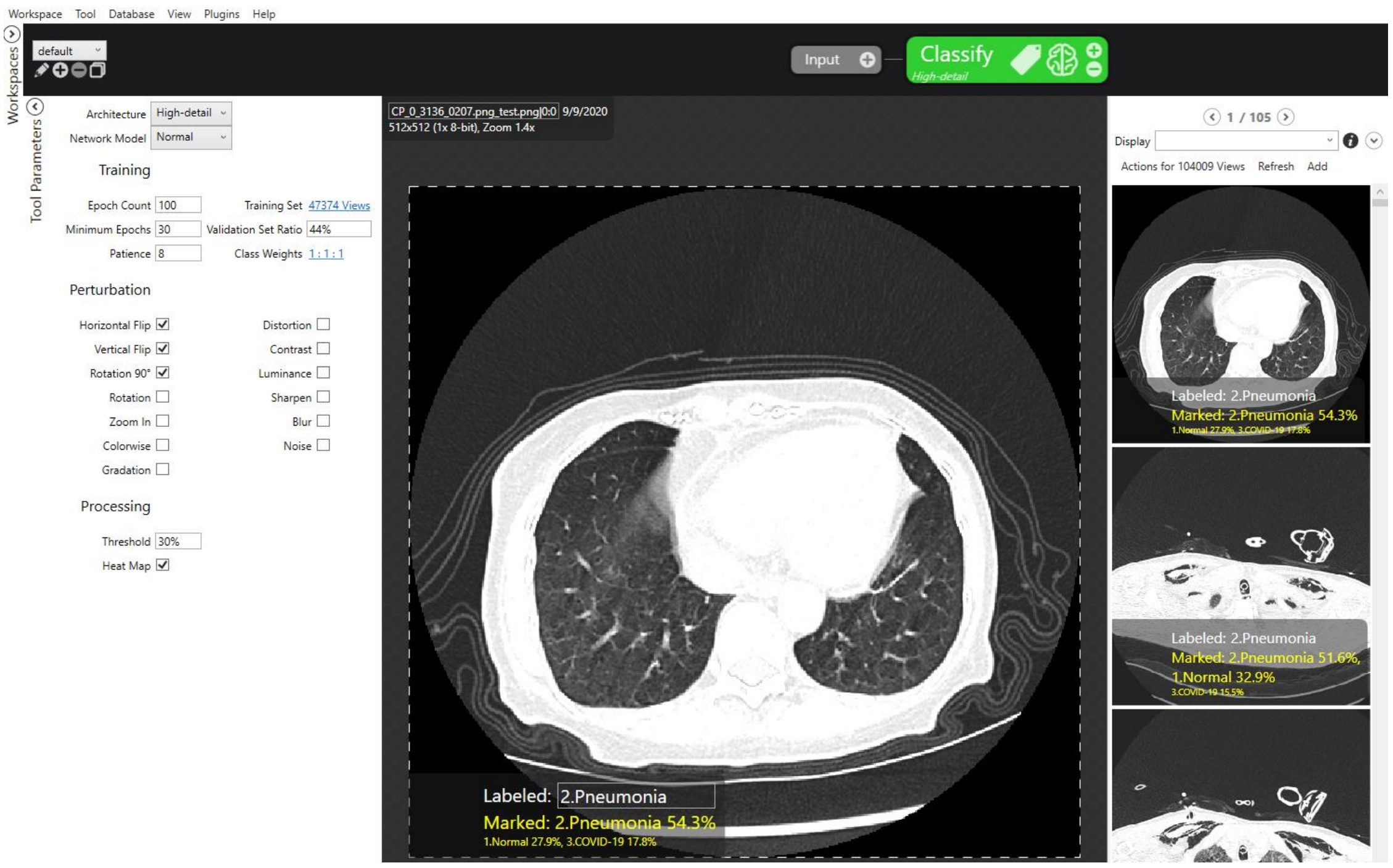

**Figure 5:** The VisionPro Deep Learning GUI loaded with the X-ray images from the COVIDx-CT dataset [18]. On the left of the GUI, there are options to select various parameters for training the model, such as model type, model size, epoch count, minimum epochs and patience, train and validation split, class weights, threshold, heat-map and the different data augmentation options of flip, rotation, contrast, zoom, brightness, sharpen, blur, distortion and noise. In the middle, the selected image is shown. On the right, thumbnails of all the images in the training, validation and test set are shown. On the top, there is the tool selection option. In the figure, the green tool has been selected for classification. Clicking on the 'brain' shaped icon in the green tool, starts the training of the model.

The maximum number of epoch counts are selected to be 100. There are options of selecting the minimum epochs and patience for which the model will train. Once these are selected, training is started by clicking on the 'brain' icon on the green tool, as seen in Figure 5.

The VisionPro Deep Learning GUI also allows geometrical data augmentation. It is called Perturbation. In this study nod data augmentation / perturbation has been used.

**4. Classification using ResNet**

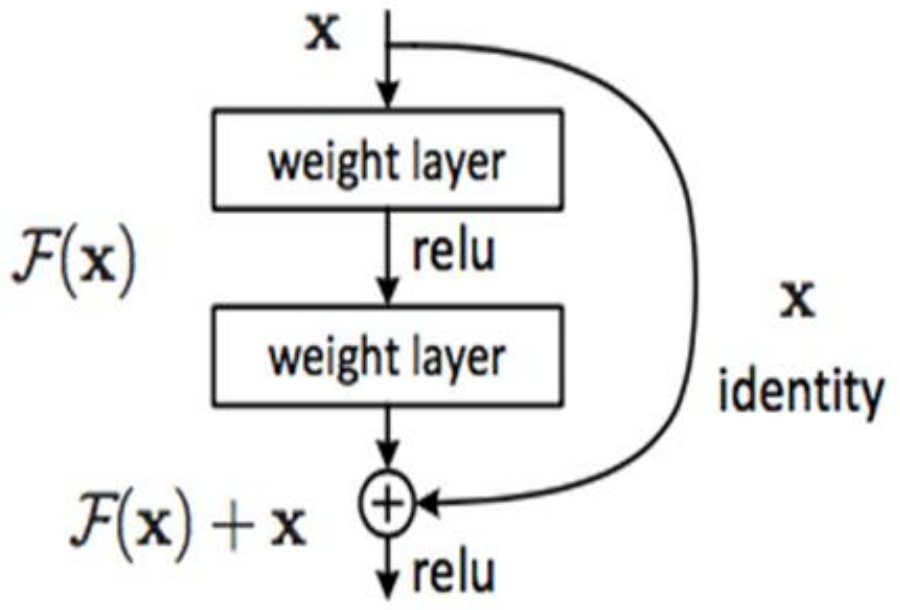

**Figure 6:** Residual learning: a building block. Image from original ResNet paper [19].

The ResNet architecture consists of several residual blocks with each block having several convolutional operations. The implementation of skip connections, as shown in Figure 5, makes the ResNet [19] better than VGG [18]. The skip connections between layers add the outputs from previous layers to the outputs of the stacked layers. This allows the training of deeper networks. One of the problems that ResNet solves is the vanishing gradient problem [25].

For training the COVIDx-CT [23] dataset we use the 50-layer ResNet50V2 (Version 2) [19] architecture. We use transfer learning to train the model, and then add two fully connected layers with L2 regularization followed by Dropouts for better regularization.

**5. Classification using DenseNet**

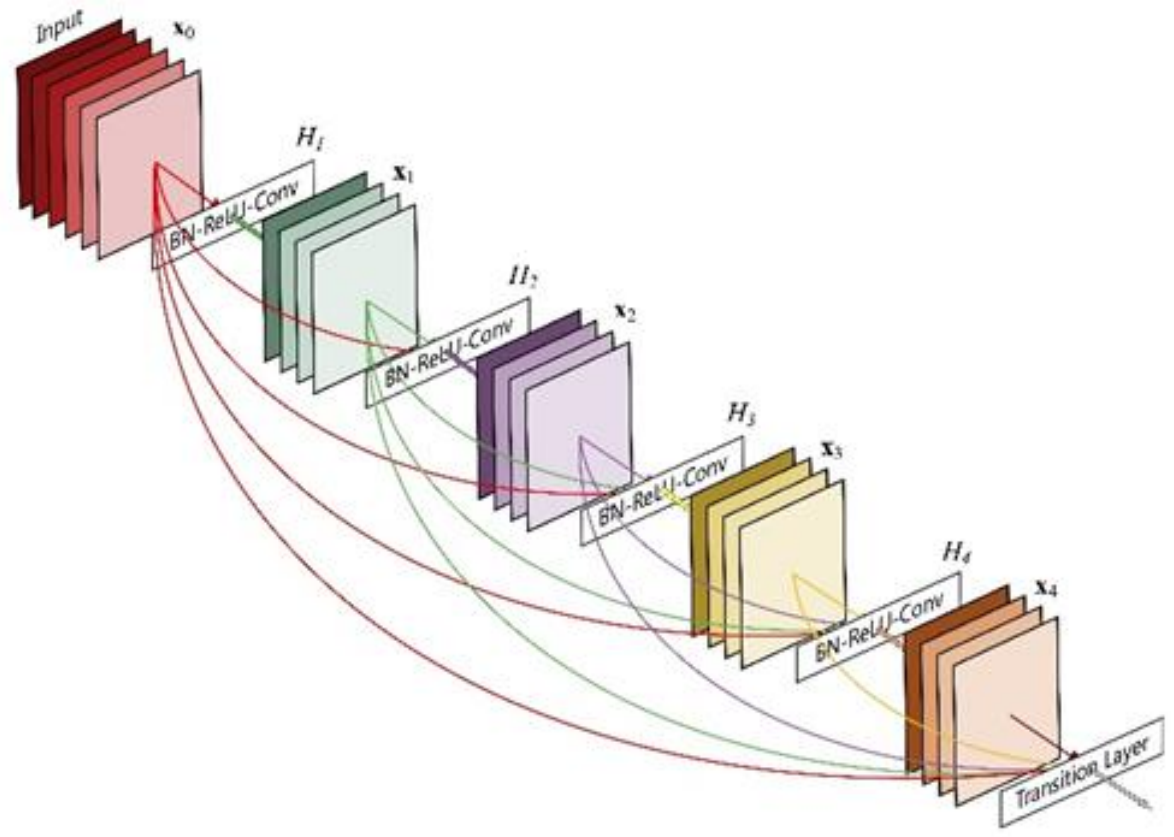

**Figure 7:** A 5-layer dense block. Each layer takes all preceding feature-maps as input. Image from original DenseNet paper [20].

DenseNet (Dense Convolutional Network) [20] is an architecture which focuses on making the Deep Learning networks go even deeper, but at the same time makes them more efficient to train, by using shorter connection between the layers. DenseNet is a convolutional neural network where each layer is connected to all other layers that are deeper in the network, that is, the first layer is connected to the $2^{nd}$, $3^{rd}$, $4^{th}$ and so on, the second layer is connected to the 3rd, 4th, 5th and so on. Unlike ResNet [19] it does not combine features through summation but combines the features by concatenating them. So, the 'ith' layer has 'i' inputs and consists of feature maps of all its preceding convolutional blocks. It hence requires fewer parameters than traditional convolutional neural networks.

For training the COVIDx-CT [23] dataset we use the 121 layered DenseNet121 [20] architecture. We use transfer learning to train the model, and then add two fully connected layers with L2 regularization followed by Dropouts for better regularization.

### 6. Classification using Inception network

Inception Network [26] has been developed with the idea of going even deeper with convolutional blocks. Very deep networks are prone to overfitting, and it is hard to pass gradient updates throughout the entire network. Also, images may have huge variations, and thus choosing the right kernel size for convolution layers is hard. To address these problems Inception network is one of the best possible networks. Inception network version 1 has multiple sizes of filters in the same level. It has various connections of 3 different sizes of filters of 1x1, 3x3, 5x5, with max pooling in a single inception module. All the outputs are concatenated and then sent to the next inception module.

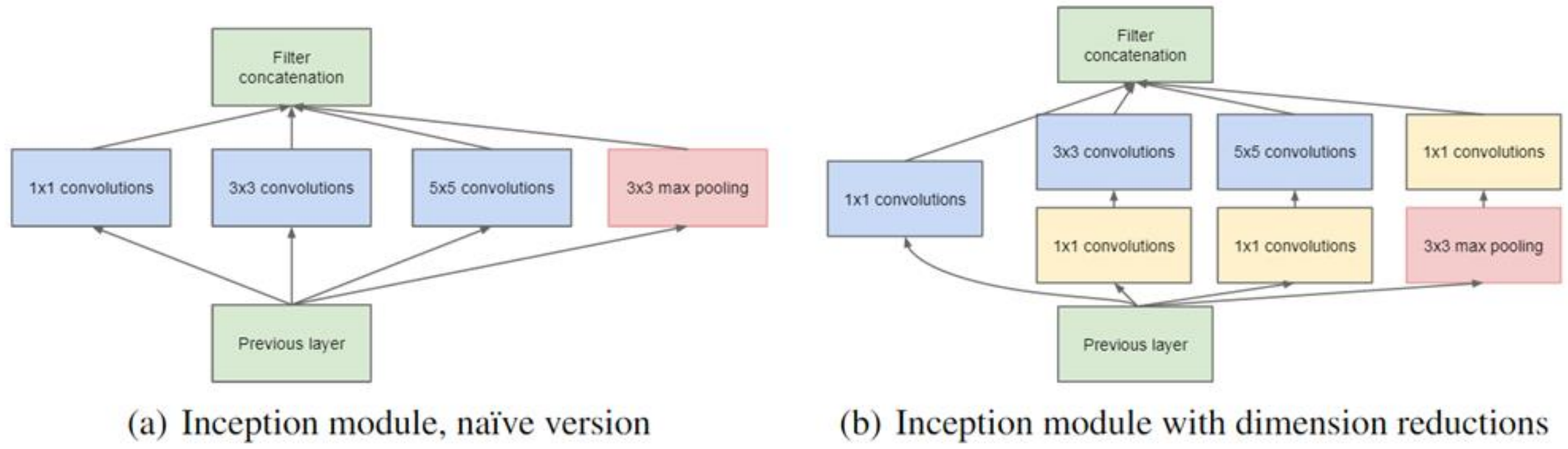

**Figure 8:** Inception Module as shown in the original Inception Version 1 paper [26].

For training the COVIDx-CT [23] dataset we use the 48 layered InceptionV3 [27] architecture, which also includes 7x7 convolutions, Batch Normalization and Label smoothing in addition to the Inception version 1 modules. We use transfer learning to train the model, and then add two fully connected layers with L2 regularization followed by Dropouts for better regularization.

### 7. Classification using Xception network

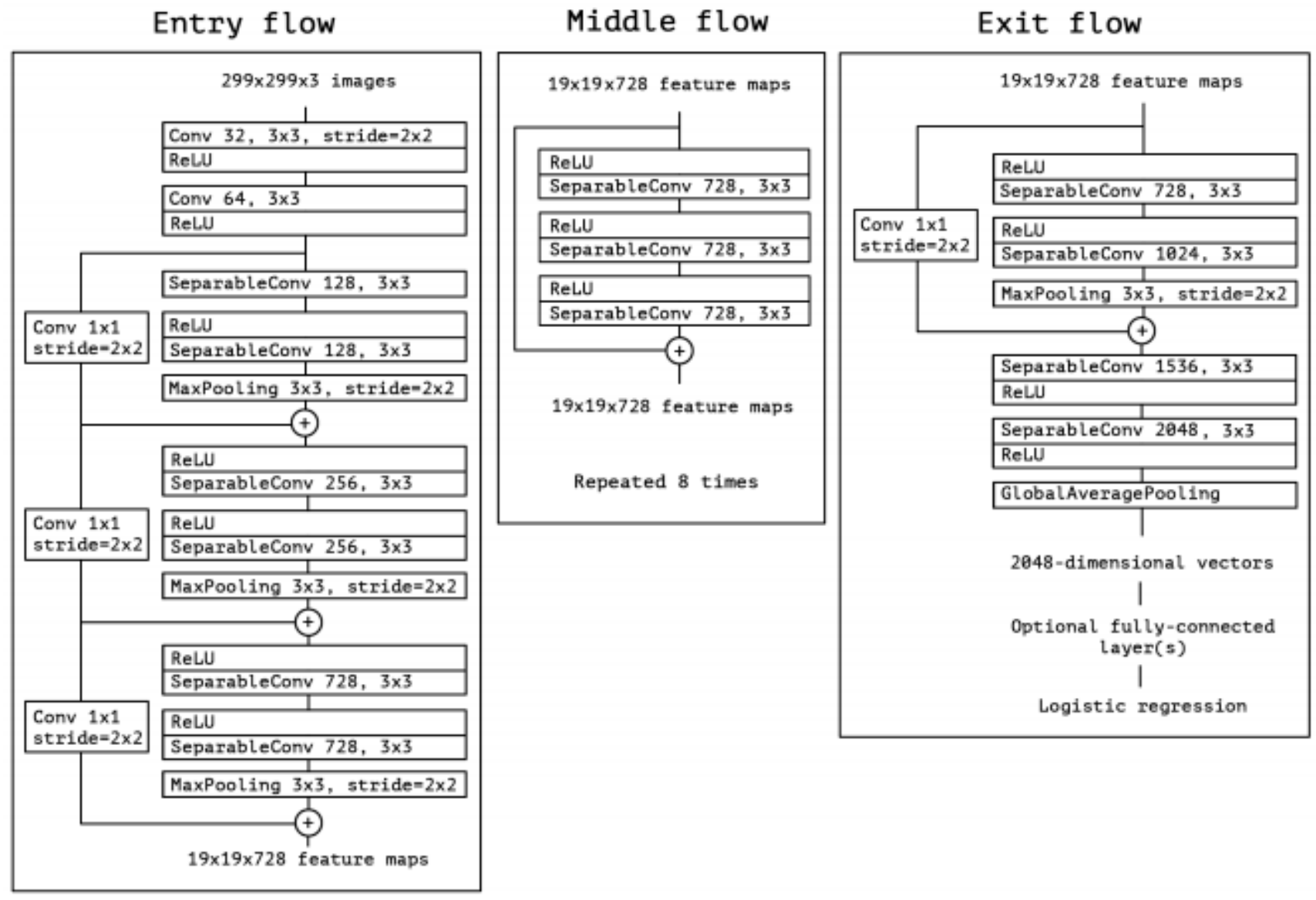

**Figure 9:** The Xception architecture. Image from the original Xception paper [28]

The Xception [28] architecture was inspired by the Inception network [26] and it employs a modified depth wise separable convolution which is a pointwise convolution followed by a depth wise convolution. The major difference in Xception is that it applies a 1x1 convolution first and then a channel wise spatial convolution, there is no intermediate

ReLU (Rectified Linear Unit) non-linearity in the modified depth wise separable convolution. This allowed the Xception [28] network to give slightly better results than the Inception network [26].

For training the COVIDx-CT [23] dataset we use the Xception [28] architecture. We use transfer learning to train the model, and then add two fully connected layers with L2 regularization followed by Dropouts for better regularization.

**RESULTS**

In medical imaging, since the decisions are of high impact, it is very important to understand exactly which evaluation metrics are necessary to decide if a model works on a patient or not. Accuracy of a model is not the best metric for deciding whether the model is fit for a patient. Rather it is important to investigate other evaluation metrices such as sensitivity, predictive values and overall F-scores.

**Sensitivity** or Recall measures the true positive rate. It is the proportion of the true positives detected by a model to the total number of positives. The better the sensitivity the better is the model at correctly identifying the infection.

$$Sensitivity = \frac{True\ Positive}{True\ Positive + False\ Negative}$$

**Positive Predictive Value (PPV)** or Precision shows the percentage of how many predictions selected the model are relevant.

$$Positive\ Predictive\ Value\ (PPV) = \frac{True\ Positive}{True\ Positive + False\ Positive}$$

**F-score** takes into consideration both the Sensitivity and PPV of a model. It can be considered as an overall score of the performance of the model.

$$F - score = \frac{True\ Positive}{True\ Positive + \left(1/2 * (False\ Positive + False\ negative)\right)}$$

A report can be generated on all the test images as seen in Figure 10. It shows a small snippet of 10 COVID-19 positive images from the test set. The report contains details of the test images, including the filename, the image, the original label as ‘Labeled’, and the predictions made by VisionPro Deep learning as ‘Marked’, with the percentage of confidence of prediction on each class.

**Confidence Interval:** A confidence interval is a range of values we are fairly sure our true value always lies in. We calculate a 95% confidence interval on the predicted sensitivity and the positive predicted values, to figure out a possible range of values by which the actual results may vary on the given test data. The confidence interval of the accuracy rates is calculated using the formula:

$$r = z\ (\surd\ (accuracy(1 - accuracy)))/N$$

where, z is the significance level of the confidence interval (the number of standard deviation of the Gaussian distribution), accuracy is the estimated accuracy (in our cases sensitivity, positive predictive value, and F-score), and N (number of images in each class) denotes the number of samples for that class. Here we use 95% confidence interval, for which the corresponding value of z is 1.96 [30].

First, the confusion matrix is plotted for the test images, for all the models that we use for the comparison. Figure 12 shows the confusion matrix of VisionPro Deep Learning. VisionPro Deep learning GUI does not display numbers of correctly classified or misclassified images on the confusion matrix. But if any point on the confusion matrix is clicked, it displays, not only the number of images in that category, but also all the images belonging to that category, with the prediction percentage and whether the prediction it made is correct or not. Below the confusion matrix it displays all the evaluation metrics of recall (sensitivity), precision (positive predictive value) and F-scores. This table contains the number of labeled images, which shows the number of images in each class in the test set. The ‘Found’ column shows the number of images VisionPro Deep Learning thinks should belong in those classes.

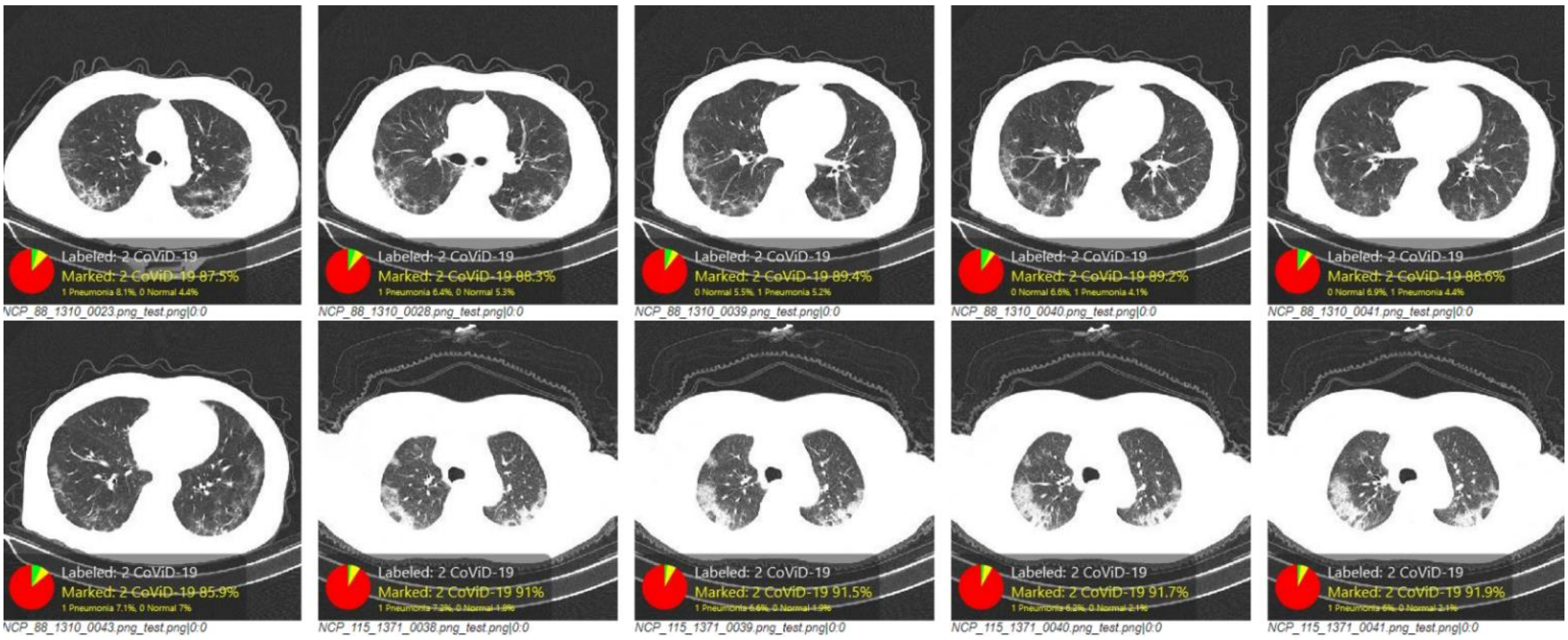

**Figure 10:** A snippet of the report generated on the test images by VisionPro Deep Learning on Setting 1. The test images are also shown with the correct labels, the predicted labels, and the confidence percentage, with pie chart, of each class. In this image, all the images are classified correctly.

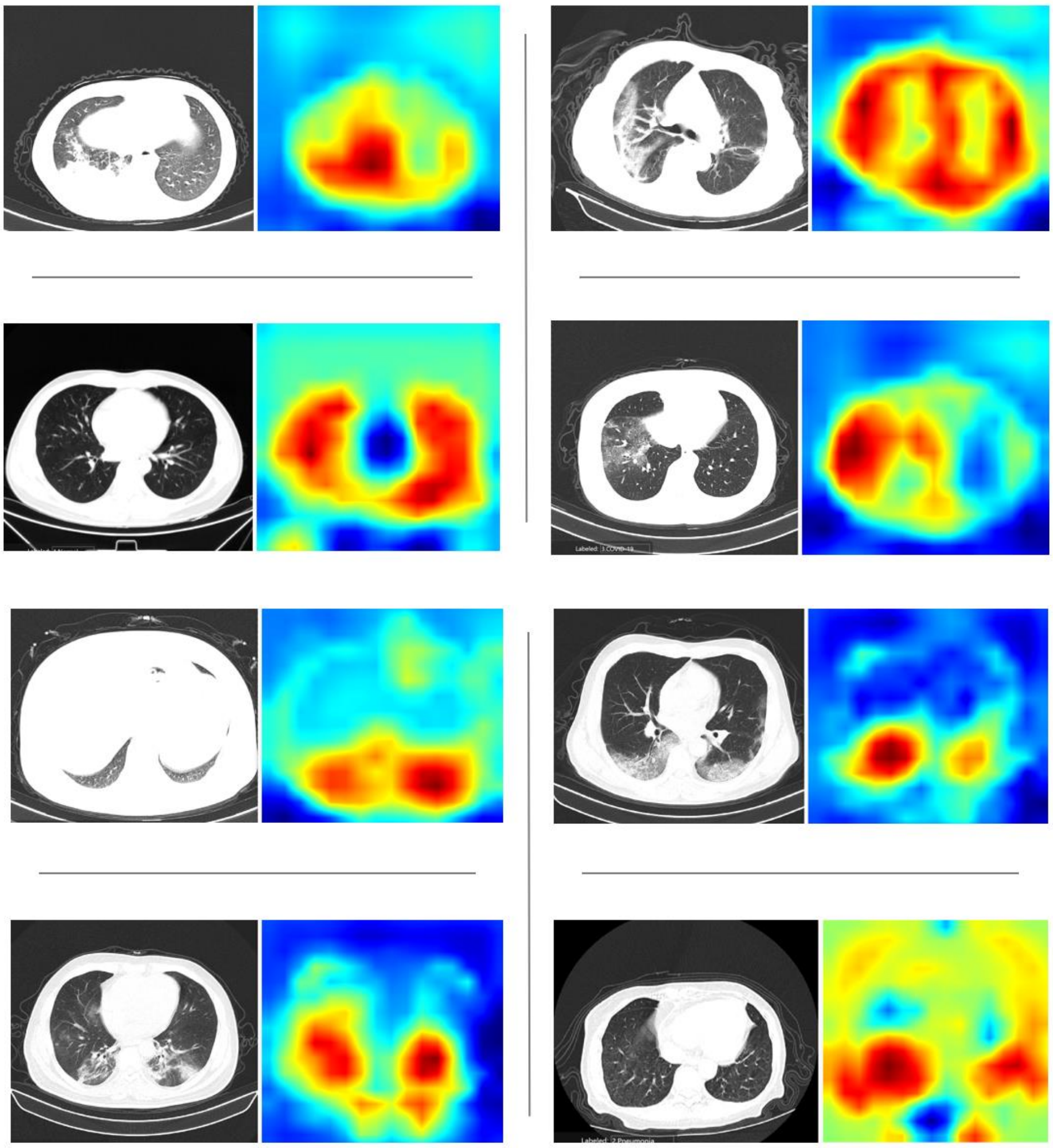

**Figure 11:** Heatmaps generated by VisionPro Deep Learning on 8 images from the test set of the COVIDx-CT images

**Setting 1:**

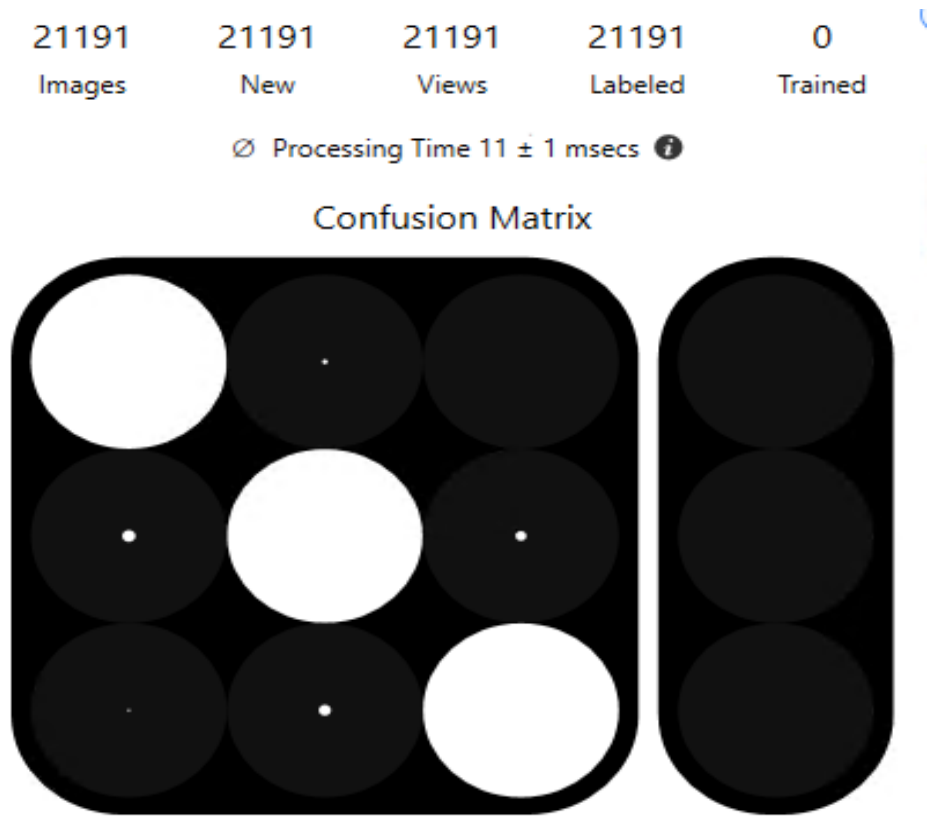

| | | | |
|---|---|---|---|
| **1 Normal** | 9442 | 8 | 0 |
| **2 Pneumonia** | 35 | 7337 | 23 |
| **3 COVID-19** | 1 | 17 | 4328 |
| | **1 Normal** | **2 Pneumonia** | **3 COVID-19** |

**Figure 12:** Confusion matrix on the 21,191 test images - COGNEX VisionPro Deep Learning with Setting 1

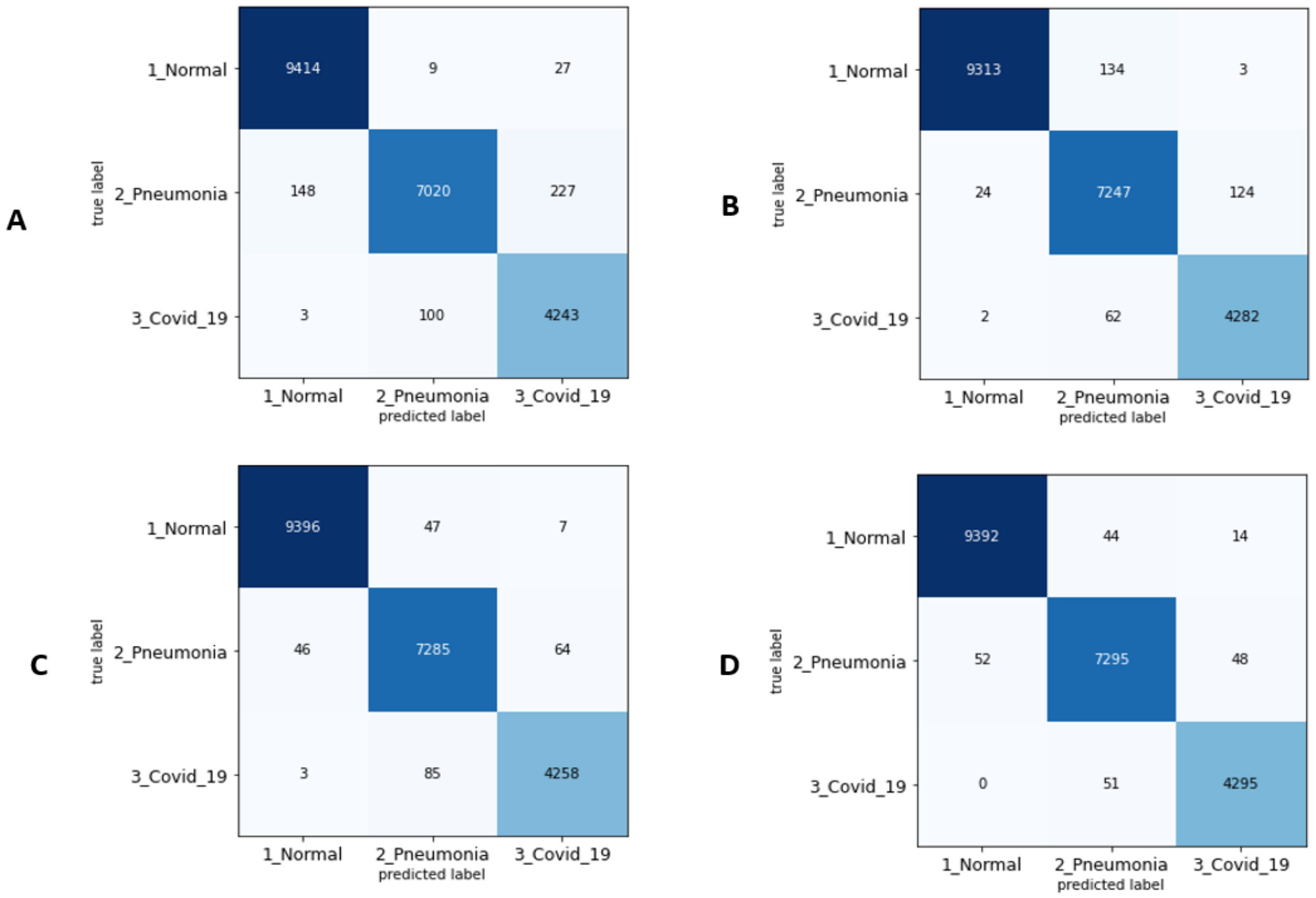

**Figure 13:** Confusion matrix on the 21,191 test images for the open-source architectures a: ResNet50 V2 [19], b: DenseNet121 [20], c: Inception V3 [27], d: Xception [28].

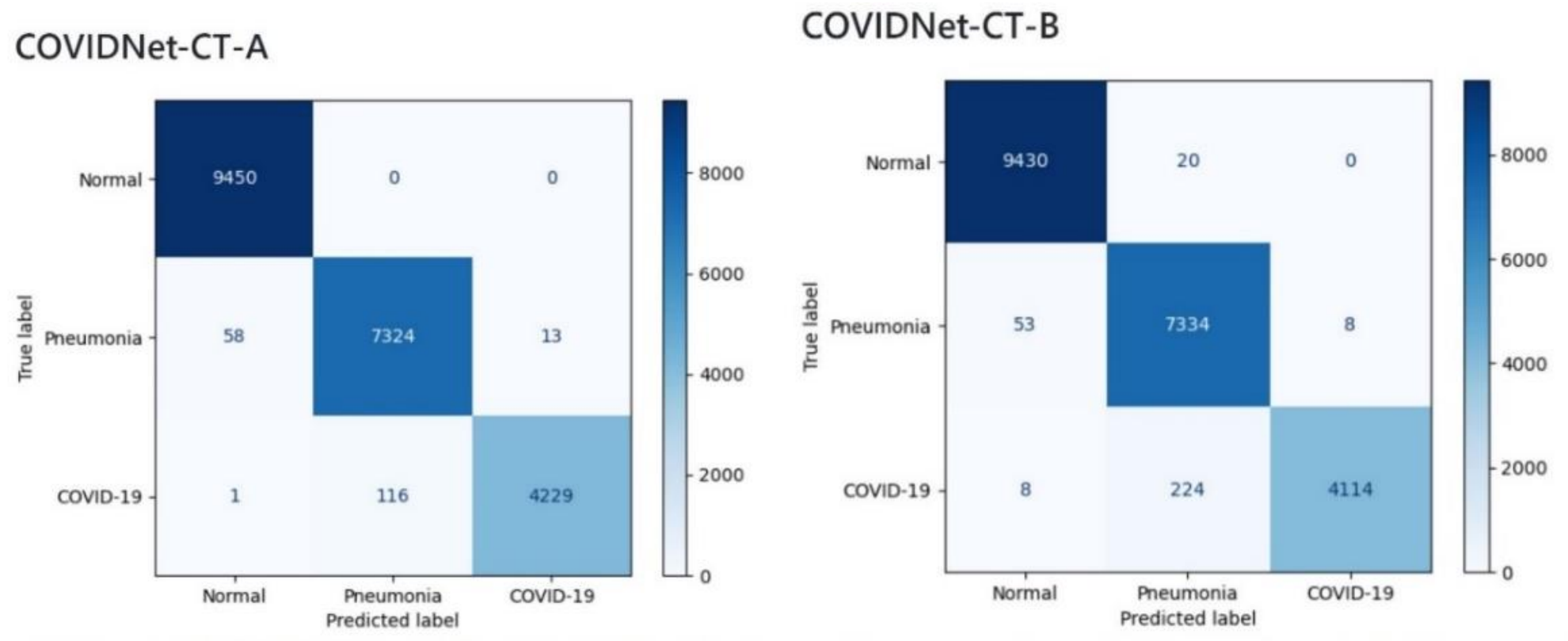

**Figure 14:** Confusion matrix on the 21,191 test images for COVIDNet-CT. Image from the original COVID-Net paper [23,29].

| Sensitivity (%) – Results Obtained | | | | |
|---|---|---|---|---|
| | | **Normal** | **Pneumonia** | **COVID-19** |
| 1 | ResNet50_v2 | 99.6 | 94.9 | 97.6 |
| 2 | DenseNet121 | 99.5 | 98.0 | 98.5 |
| 3 | Inception_v3 | 99.4 | 98.5 | 97.9 |
| 4 | Xception | 99.3 | 98.6 | 98.8 |
| 5 | COVID-Net CT A | **100.0** | 99.0 | 97.3 |
| 6 | COVID-Net CT B | 99.8 | **99.2** | 94.7 |
| 7 | COGNEX VisionPro Deep Learning 1.0 | 99.9 | **99.2** | **99.5** |

**Table 4a.1:** Sensitivity for each infection type. Best results are highlighted in **BOLD**

| Sensitivity (%) – 95% Confidence Interval | | | | |
|---|---|---|---|---|
| | | **Normal** | **Pneumonia** | **COVID-19** |
| 1 | ResNet50_v2 | 99.6 ± 0.12 | 94.9 ± 0.50 | 97.6 ± 0.45 |
| 2 | DenseNet121 | 99.5 ± 0.14 | 98.0 ± 0.31 | 98.5 ± 0.36 |
| 3 | Inception_v3 | 99.4 ± 0.15 | 98.5 ± 0.27 | 97.9 ± 0.42 |
| 4 | Xception | 99.3 ± 0.16 | 98.6 ± 0.26 | 98.8 ± 0.32 |
| 5 | COVID-Net CT A | **100.0 ± 0.00** | 99.0 ± 0.22 | 97.3 ± 0.48 |
| 6 | COVID-Net CT B | 99.8 ± 0.09 | **99.2 ± 0.20** | 94.7 ± 0.66 |
| 7 | COGNEX VisionPro Deep Learning 1.0 | 99.9 ± 0.06 | **99.2 ± 0.20** | **99.5 ± 0.20** |

**Table 4a.2**: Sensitivity calculated with 95% Confidence Interval. Best results are highlighted in **BOLD**

| PPV (%) – Results Obtained | | | | |
|---|---|---|---|---|
| | | **Normal** | **Pneumonia** | **COVID-19** |
| 1 | ResNet50_v2 | 98.4 | 98.4 | 94.3 |
| 2 | DenseNet121 | **99.7** | 97.3 | 97.1 |
| 3 | Inception_v3 | 99.4 | 98.2 | 98.3 |
| 4 | Xception | 99.4 | 98.7 | 98.5 |
| 5 | COVID-Net CT A | 99.4 | 98.4 | 99.7 |
| 6 | COVID-Net CT B | 99.4 | 96.8 | **99.8** |
| 7 | COGNEX VisionPro Deep Learning 1.0 | 99.6 | **99.6** | 99.4 |

**Table 4b.1:** PPV for each infection type. Best results are highlighted in **BOLD**

| PPV (%) – 95% Confidence Interval | | | | |
|---|---|---|---|---|
| | | **Normal** | **Pneumonia** | **COVID-19** |
| 1 | ResNet50_v2 | 98.4 ± 0.25 | 98.4 ± 0.28 | 94.3 ± 0.68 |
| 2 | DenseNet121 | **99.7 ± 0.11** | 97.3 ± 0.36 | 97.1 ± 0.49 |
| 3 | Inception_v3 | 99.4 ± 0.15 | 98.2 ± 0.30 | 98.3 ± 0.38 |
| 4 | Xception | 99.4 ± 0.15 | 98.7 ± 0.25 | 98.5 ± 0.36 |
| 5 | COVID-Net CT A | 99.4 ± 0.15 | 98.4 ± 0.28 | 99.7 ± 0.16 |
| 6 | COVID-Net CT B | 99.4 ± 0.15 | 96.8 ± 0.40 | **99.8 ± 0.13** |
| 7 | COGNEX VisionPro Deep Learning 1.0 | 99.6 ± 0.12 | **99.6 ± 0.14** | 99.4 ± 0.22 |

**Table 4b.2:** PPV calculated with 95% Confidence Interval. Best results are highlighted in **BOLD**

| F-score (%) – Results Obtained | | | | |
|---|---|---|---|---|
| | | **Normal** | **Pneumonia** | **COVID-19** |
| 1 | ResNet50_v2 | 99.0 | 96.6 | 95.9 |
| 2 | DenseNet121 | 99.1 | 97.6 | 97.8 |
| 3 | Inception_v3 | 99.4 | 98.3 | 98.1 |
| 4 | Xception | 99.4 | 98.6 | 98.7 |
| 5 | COVID-Net CT A | 99.6 | 98.6 | 98.4 |
| 6 | COVID-Net CT B | 99.6 | 97.9 | 97.1 |
| 7 | COGNEX VisionPro Deep Learning 1.0 | **99.7** | **99.4** | **99.5** |

**Table 4c.1:** F-score for each infection type. Best results are highlighted in **BOLD**

| **F-score (%) – 95% Confidence Interval** | | | | |
|---|---|---|---|---|
| | | **Normal** | **Pneumonia** | **COVID-19** |
| 1 | ResNet50_v2 | 99.0 ± 0.20 | 96.6 ± 0.41 | 95.9 ± 0.58 |
| 2 | DenseNet121 | 99.1 ± 0.19 | 97.6 ± 0.34 | 97.8 ± 0.43 |
| 3 | Inception_v3 | 99.4 ± 0.15 | 98.3 ± 0.29 | 98.1 ± 0.40 |
| 4 | Xception | 99.4 ± 0.15 | 98.6 ± 0.26 | 98.7 ± 0.33 |
| 5 | COVID-Net CT A | 99.6 ± 0.12 | 98.6 ± 0.26 | 98.4 ± 0.37 |
| 6 | COVID-Net CT B | 99.6 ± 0.12 | 97.9 ± 0.32 | 97.1 ± 0.49 |
| 7 | COGNEX VisionPro Deep Learning 1.0 | **99.7 ± 0.11** | **99.4 ± 0.17** | **99.5 ± 0.20** |

**Table 4c.2:** F-score calculated with 95% Confidence Interval. Best results are highlighted in **BOLD**

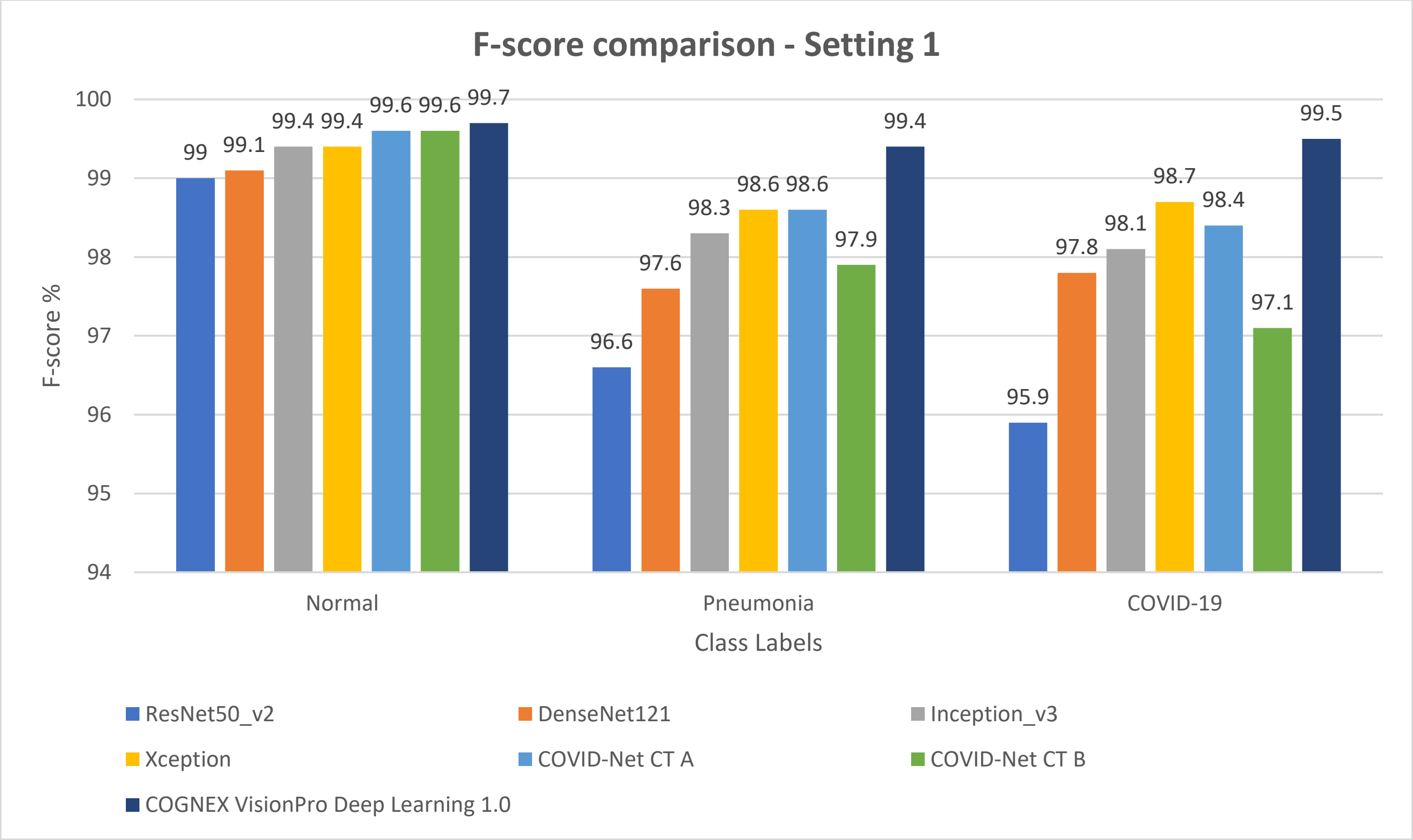

**Figure 15:** F-score comparison – Bar Chart – Setting 1

In Sensitivity, for Normal cases COVID-Net CT A [23][29] gets a score of 100%. For images of Pneumonia VisionPro Deep Learning and COVID-Net CT B [23][29] have the highest scores. For the COVID-19 images, VisionPro Deep Learning has the highest score of 99.5%, which is quite a way ahead of the next best result by Xception [28]. For PPV, DenseNet [20] has the best scores for Normal cases. VisionPro Deep Learning for Pneumonia, and COVID-Net CT B [23][29] for COVID-19 images. For F-scores, VisionPro Deep Learning has the best results for all the three classes with a score of 99.5% for COVID-19 images. Xception [28] performs really well, with scores even better than COVID-Net CT [23] for the COVID-19 class.

**Setting 2:**

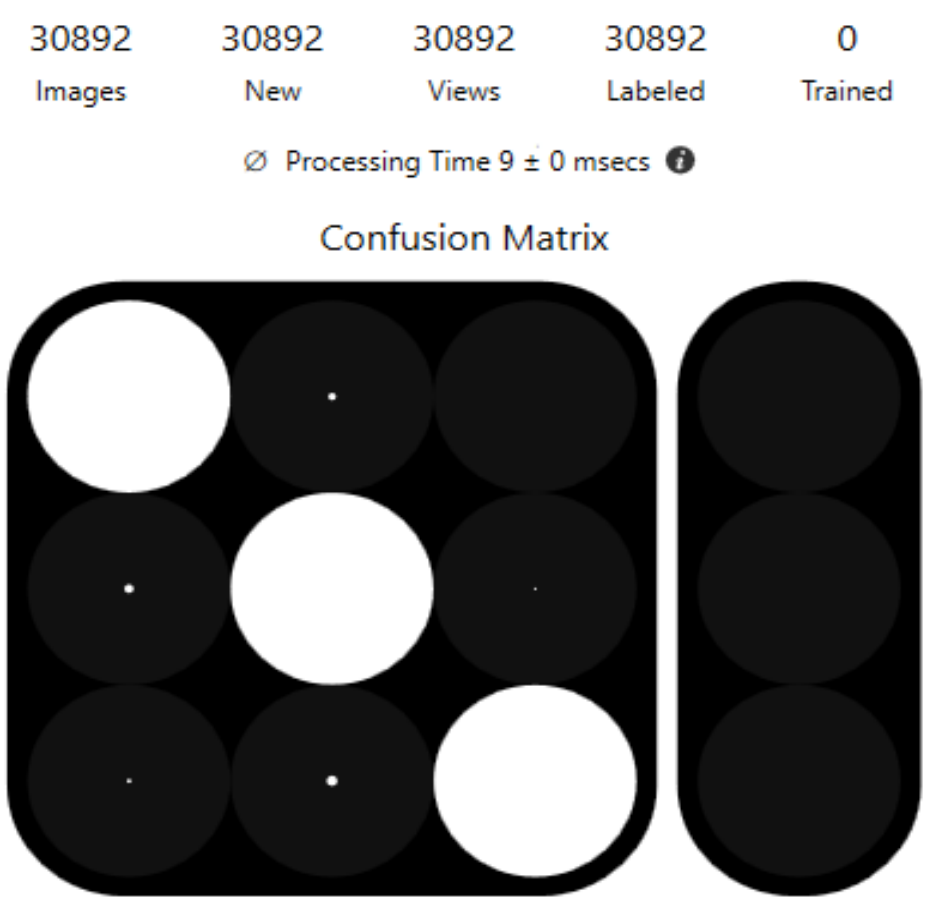

| | | | |
|---|---|---|---|
| **1 Normal** | 12650 | 20 | 0 |
| **2 Pneumonia** | 24 | 10651 | 2 |
| **3 COVID-19** | 4 | 22 | 7519 |
| | **1 Normal** | **2 Pneumonia** | **3 COVID-19** |

**Figure 16:** Confusion matrix on the 30,892 test images - COGNEX VisionPro Deep Learning with Setting 2

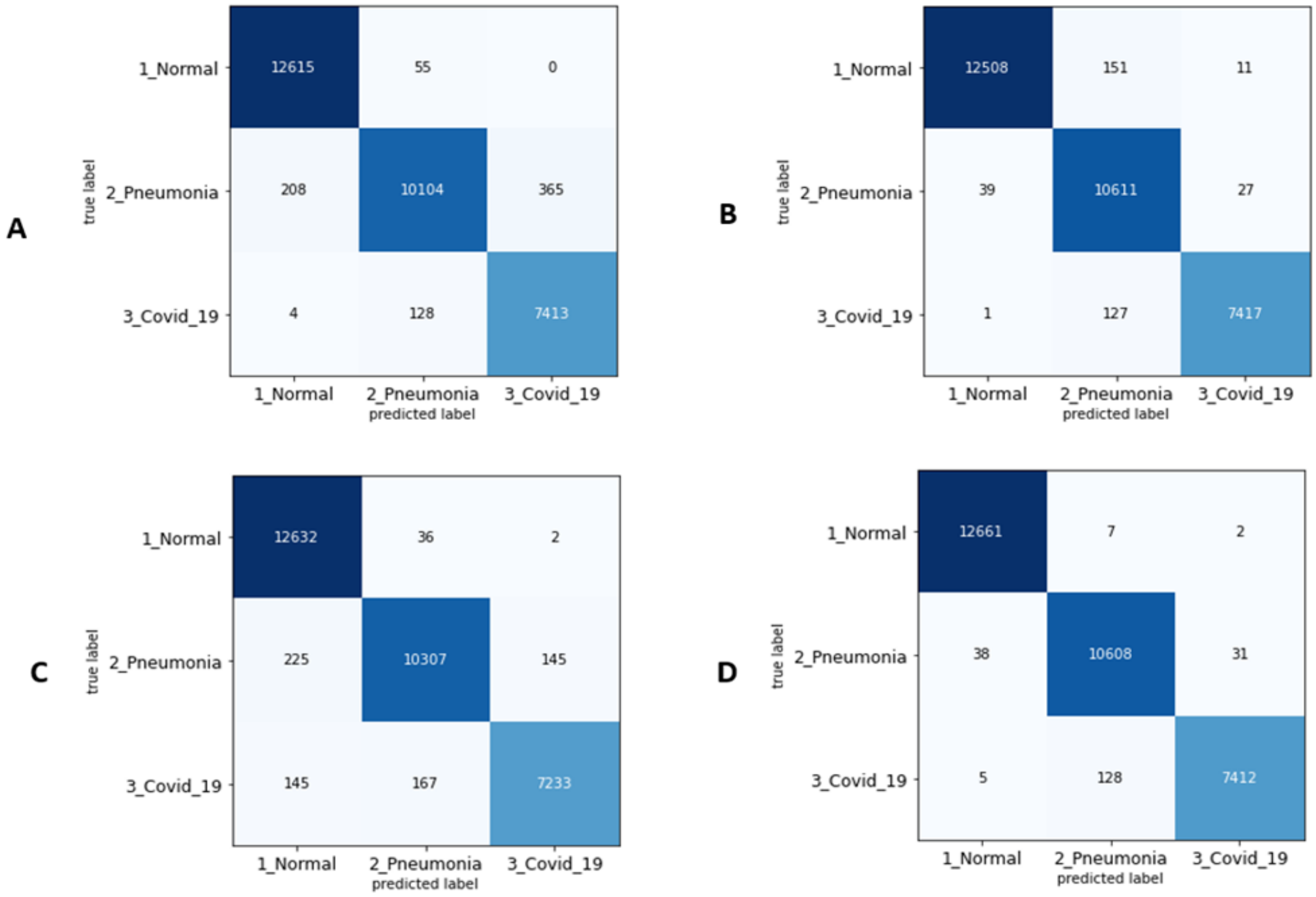

**Figure 17:** Confusion matrix on the 30,892 test images for the open-source architectures a: ResNet50 V2 [19], b: DenseNet121 [20], c: Inception V3 [27], d: Xception [28].

| **Sensitivity (%) – Results Obtained** | | | | |
|---|---|---|---|---|
| | | **Normal** | **Pneumonia** | **COVID-19** |
| 1 | ResNet50_v2 | 99.5 | 94.6 | 98.2 |
| 2 | DenseNet121 | 98.7 | 99.3 | 98.3 |
| 3 | Inception_v3 | 99.7 | 96.5 | 95.8 |
| 4 | Xception | **99.9** | 99.3 | 98.2 |
| 5 | COGNEX VisionPro Deep Learning 1.0 | 99.8 | **99.7** | **99.6** |

**Table 5a.1:** Sensitivity for each infection type. Best results are highlighted in **BOLD**

| Sensitivity (%) – 95% Confidence Interval | | | | |
|---|---|---|---|---|
| | | **Normal** | **Pneumonia** | **COVID-19** |
| 1 | ResNet50_v2 | 99.5 ± 0.12 | 94.6 ± 0.42 | 98.2 ± 0.29 |
| 2 | DenseNet121 | 98.7 ± 0.19 | 99.3 ± 0.15 | 98.3 ± 0.29 |
| 3 | Inception_v3 | 99.7 ± 0.09 | 96.5 ± 0.34 | 95.8 ± 0.45 |
| 4 | Xception | **99.9 ± 0.05** | 99.3 ± 0.15 | 98.2 ± 0.29 |
| 5 | COGNEX VisionPro Deep Learning 1.0 | 99.8 ± 0.07 | **99.7 ± 0.10** | **99.6 ± 0.14** |

**Table 5a.2:** Sensitivity calculated with 95% Confidence Interval. Best results are highlighted in **BOLD**

| PPV (%) – Results Obtained | | | | |
|---|---|---|---|---|
| | | **Normal** | **Pneumonia** | **COVID-19** |
| 1 | ResNet50_v2 | 98.3 | 98.2 | 95.3 |
| 2 | DenseNet121 | 99.6 | 97.4 | 99.4 |
| 3 | Inception_v3 | 97.1 | 98.0 | 98.0 |
| 4 | Xception | 99.6 | 98.7 | 99.5 |
| 5 | COGNEX VisionPro Deep Learning 1.0 | **99.7** | **99.6** | **99.9** |

**Table 5b.1:** PPV for each infection type. Best results are highlighted in **BOLD**

| PPV (%) – 95% Confidence Interval | | | | |
|---|---|---|---|---|
| | | **Normal** | **Pneumonia** | **COVID-19** |
| 1 | ResNet50_v2 | 98.3 ± 0.22 | 98.2 ± 0.25 | 95.3 ± 0.47 |
| 2 | DenseNet121 | 99.6 ± 0.10 | 97.4 ± 0.30 | 99.4 ± 0.17 |
| 3 | Inception_v3 | 97.1 ± 0.29 | 98.0 ± 0.26 | 98.0 ± 0.31 |
| 4 | Xception | 99.6 ± 0.10 | 98.7 ± 0.21 | 99.5 ± 0.15 |
| 5 | COGNEX VisionPro Deep Learning 1.0 | **99.7 ± 0.09** | **99.6 ± 0.11** | **99.9 ± 0.07** |

**Table 5b.2:** PPV calculated with 95% Confidence Interval. Best results are highlighted in **BOLD**

| F-score (%) – Results Obtained | | | | |
|---|---|---|---|---|
| | | **Normal** | **Pneumonia** | **COVID-19** |
| 1 | ResNet50_v2 | 98.9 | 96.3 | 96.7 |
| 2 | DenseNet121 | 99.2 | 98.4 | 98.8 |
| 3 | Inception_v3 | 98.4 | 97.3 | 96.9 |
| 4 | Xception | **99.8** | 99.0 | 98.8 |
| 5 | COGNEX VisionPro Deep Learning 1.0 | **99.8** | **99.6** | **99.8** |

**Table 5c.1:** F-score for each infection type. Best results are highlighted in **BOLD**

| F-score (%) – 95% Confidence Interval | | | | |
|---|---|---|---|---|
| | | **Normal** | **Pneumonia** | **COVID-19** |
| 1 | ResNet50_v2 | 98.9 ± 0.18 | 96.3 ± 0.35 | 96.7 ± 0.40 |
| 2 | DenseNet121 | 99.2 ± 0.15 | 98.4 ± 0.23 | 98.8 ± 0.24 |
| 3 | Inception_v3 | 98.4 ± 0.21 | 97.3 ± 0.30 | 96.9 ± 0.39 |
| 4 | Xception | **99.8 ± 0.07** | 99.0 ± 0.18 | 98.8 ± 0.24 |
| 5 | COGNEX VisionPro Deep Learning 1.0 | **99.8 ± 0.07** | **99.6 ± 0.11** | **99.8 ± 0.10** |

**Table 5c.2:** F-score calculated with 95% Confidence Interval. Best results are highlighted in **BOLD**

Setting 2 has an approximate of 31,000 images in the test set. In this setting, since we do not have COVID-Net CT [23] results, we use the other open-source architectures for comparison.

In Sensitivity, for Normal cases, Xception [28] gives the best result, but for images of Pneumonia and COVID-19, VisionPro Deep Learning has the best results with scores above 99.5% in both cases.

In PPV, VisionPro Deep learning has the best results for all the three classes, with all the scores above 99.5%.

In F-score, for Normal cases, Xception [28] and VisionPro Deep Learning have the same scores of 99.8%, but for Pneumonia and COVID-19 cases, VisionPro Deep learning has the highest scores of 99.6% and 99.8%, respectively.

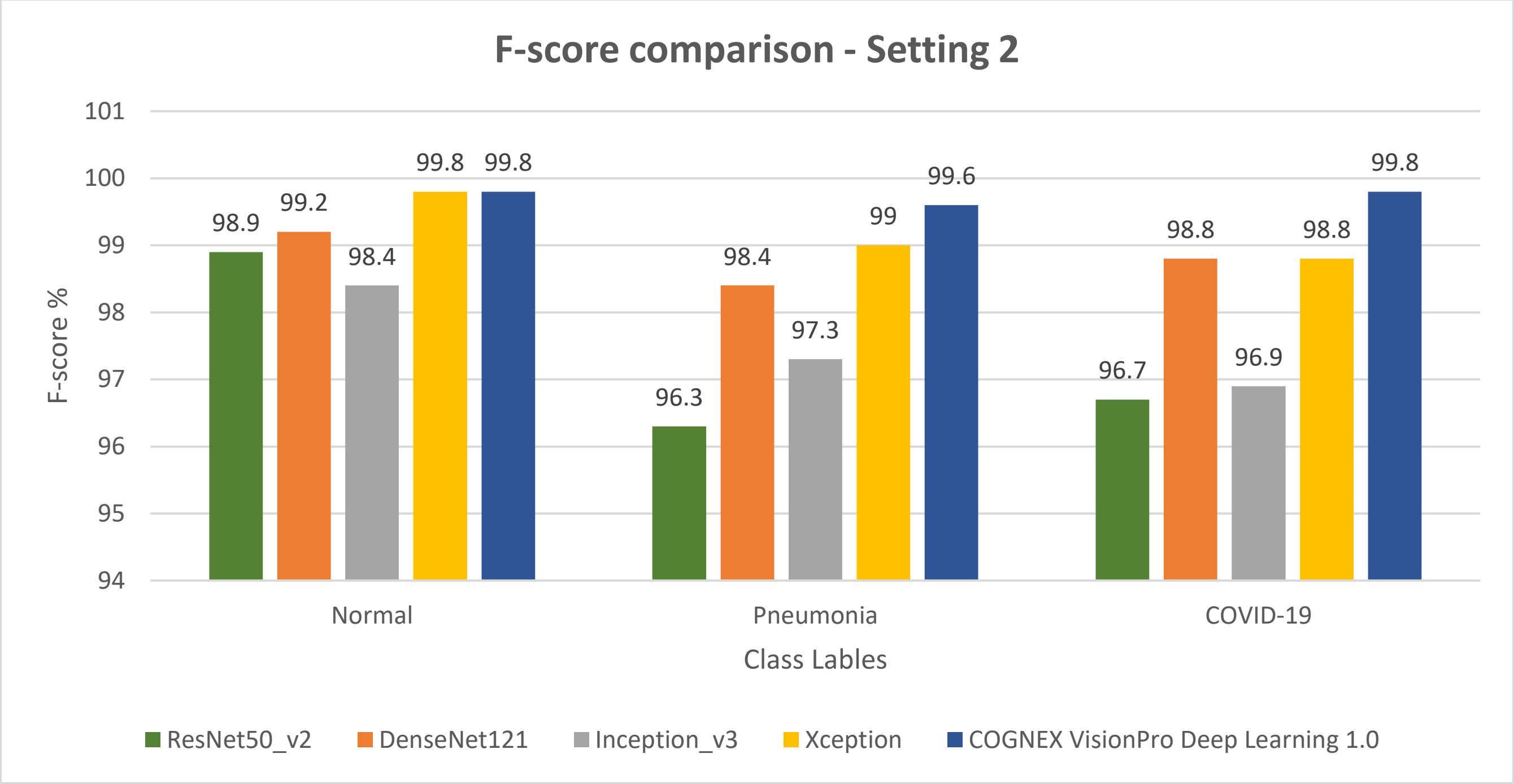

**Figure 18:** F-score comparison – Bar Chart – Setting 2

**Setting 3:**

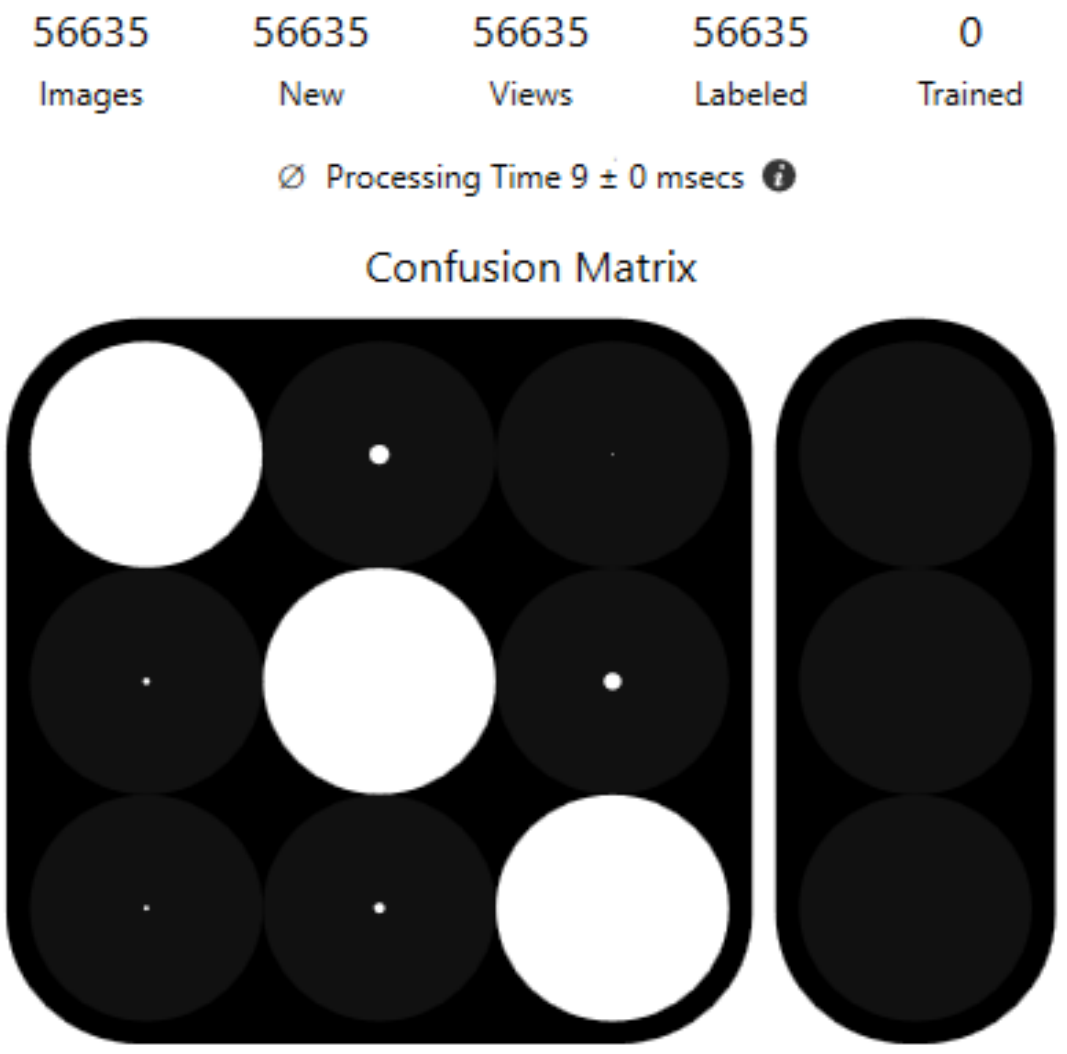

| | | | |
|---|---|---|---|
| **1 Normal** | 24003 | 181 | 3 |
| **2 Pneumonia** | 20 | 20082 | 116 |
| **3 COVID-19** | 7 | 28 | 12195 |
| | **1 Normal** | **2 Pneumonia** | **3 COVID-19** |

**Figure 18:** Confusion matrix on the 56,635 test images - COGNEX VisionPro Deep Learning with Setting 3

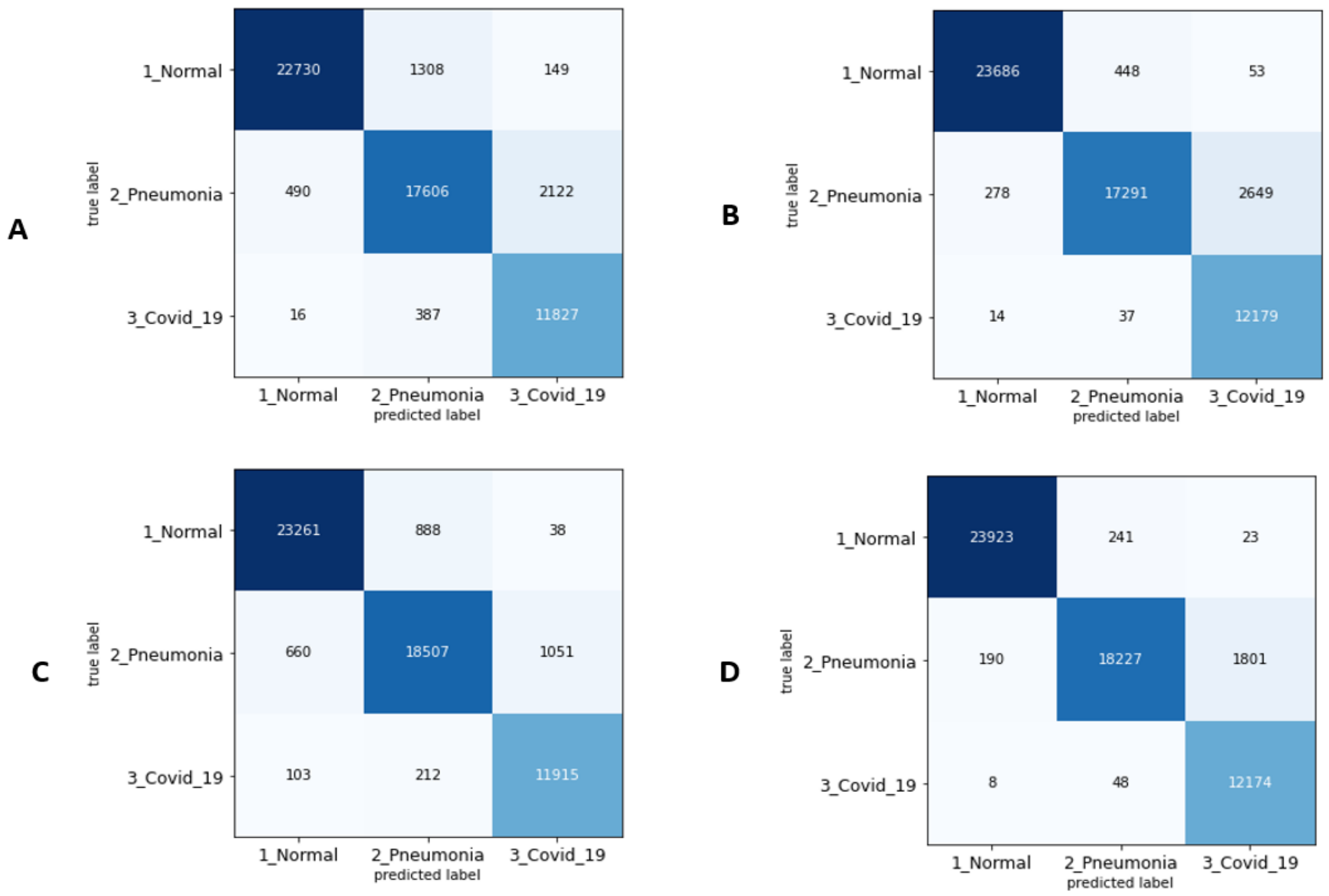

**Figure 19:** Confusion matrix on the 56,635 test images for the open-source architectures a: ResNet50 V2 [19], b: DenseNet121 [20], c: Inception V3 [27], d: Xception [28].

| Sensitivity (%) – Results Obtained | | | | |
|---|---|---|---|---|
| | | **Normal** | **Pneumonia** | **COVID-19** |
| 1 | ResNet50_v2 | 93.9 | 87.0 | 96.7 |
| 2 | DenseNet121 | 97.9 | 85.5 | 99.5 |
| 3 | Inception_v3 | 96.1 | 91.5 | 97.4 |
| 4 | Xception | 98.9 | 90.1 | 99.5 |
| 5 | COGNEX VisionPro Deep Learning 1.0 | **99.2** | **99.3** | **99.7** |

**Table 6a.1:** Sensitivity for each infection type. Best results are highlighted in **BOLD**

| Sensitivity (%) – 95% Confidence Interval | | | | |
|---|---|---|---|---|
| | | **Normal** | **Pneumonia** | **COVID-19** |
| 1 | ResNet50_v2 | 93.9 ± 0.30 | 87.0 ± 0.46 | 96.7 ± 0.31 |
| 2 | DenseNet121 | 97.9 ± 0.18 | 85.5 ± 0.48 | 99.5 ± 0.12 |
| 3 | Inception_v3 | 96.1 ± 0.24 | 91.5 ± 0.38 | 97.4 ± 0.28 |
| 4 | Xception | 98.9 ± 0.13 | 90.1 ± 0.41 | 99.5 ± 0.12 |
| 5 | COGNEX VisionPro Deep Learning 1.0 | **99.2 ± 0.11** | **99.3 ± 0.11** | **99.7 ± 0.09** |

**Table 6a.2:** Sensitivity calculated with 95% Confidence Interval. Best results are highlighted in **BOLD**

| PPV (%) – Results Obtained | | | | |
|---|---|---|---|---|
| | | **Normal** | **Pneumonia** | **COVID-19** |
| 1 | ResNet50_v2 | 97.8 | 91.2 | 83.8 |
| 2 | DenseNet121 | 98.7 | 97.2 | 81.8 |
| 3 | Inception_v3 | 96.8 | 94.3 | 91.6 |
| 4 | Xception | 99.1 | 98.4 | 86.9 |
| 5 | COGNEX VisionPro Deep Learning 1.0 | **99.8** | **99.0** | **99.0** |

**Table 6b.1:** PPV for each infection type. Best results are highlighted in **BOLD**

| PPV (%) – 95% Confidence Interval | | | | |
|---|---|---|---|---|
| | | **Normal** | **Pneumonia** | **COVID-19** |
| 1 | ResNet50_v2 | 97.8 ± 0.18 | 91.2 ± 0.39 | 83.8 ± 0.65 |
| 2 | DenseNet121 | 98.7 ± 0.14 | 97.2 ± 0.22 | 81.8 ± 0.68 |
| 3 | Inception_v3 | 96.8 ± 0.22 | 94.3 ± 0.31 | 91.6 ± 0.49 |
| 4 | Xception | 99.1 ± 0.11 | 98.4 ± 0.17 | 86.9 ± 0.59 |
| 5 | COGNEX VisionPro Deep Learning 1.0 | **99.8 ± 0.05** | **99.0 ± 0.13** | **99.0 ± 0.17** |

**Table 6b.2:** PPV calculated with 95% Confidence Interval. Best results are highlighted in **BOLD**

| F-score (%) – Results Obtained | | | | |
|---|---|---|---|---|
| | | **Normal** | **Pneumonia** | **COVID-19** |
| 1 | ResNet50_v2 | 95.8 | 89.1 | 89.8 |
| 2 | DenseNet121 | 98.3 | 91.0 | 89.8 |
| 3 | Inception_v3 | 96.5 | 92.9 | 94.4 |
| 4 | Xception | 99.0 | 94.1 | 92.8 |
| 5 | COGNEX VisionPro Deep Learning 1.0 | **99.5** | **99.1** | **99.3** |

**Table 6c.1:** F-score for each infection type. Best results are highlighted in **BOLD**

| F-score (%) – 95% Confidence Interval | | | | |
|---|---|---|---|---|
| | | **Normal** | **Pneumonia** | **COVID-19** |
| 1 | ResNet50_v2 | 95.8 ± 0.25 | 89.1 ± 0.42 | 89.8 ± 0.53 |
| 2 | DenseNet121 | 98.3 ± 0.16 | 91.0 ± 0.39 | 89.8 ± 0.53 |
| 3 | Inception_v3 | 96.5 ± 0.23 | 92.9 ± 0.35 | 94.4 ± 0.40 |
| 4 | Xception | 99.0 ± 0.12 | 94.1 ± 0.32 | 92.8 ± 0.45 |
| 5 | COGNEX VisionPro Deep Learning 1.0 | **99.5 ± 0.08** | **99.1 ± 0.13** | **99.3 ± 0.14** |

**Table 6c.2:** F-score calculated with 95% Confidence Interval. Best results are highlighted in **BOLD**

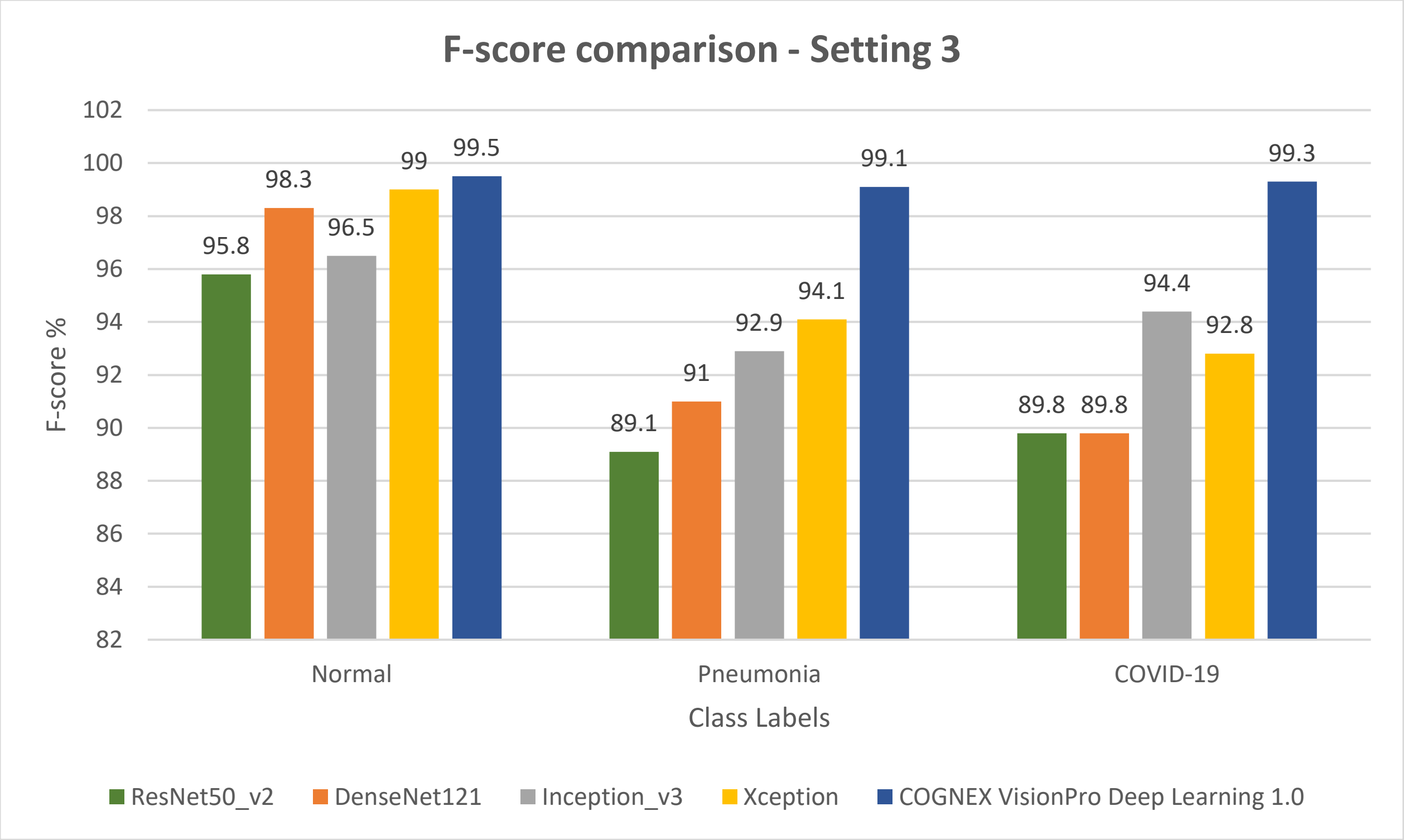

**Figure 20:** F-score comparison – Bar Chart – Setting 3

Setting 3 has an approximate of 56,000 images. That is around 35,000 more images than the original COVIDx-CT dataset. The number of training images is significantly reduced. With the lesser number of images in the training set, the performance of the open-source models decreases significantly.

VisionPro Deep Learning gets the best results for all three classes, for Sensitivity, PPV as well as F-scores. For Sensitivity, Xception [28] performs well too. In PPV, for the COVID-19 class, VisionPro Deep Learning performs significantly better than all other open-source models. VisionPro Deep Learning has a PPV of 99% for the COVID-19 class, while the next best result is from Inception network [27], with a score of 91.6%. In F-scores, for the Pneumonia and the COVID-19 class, VisionPro Deep Learning has really good results as compared to the other open-source models.

## DISCUSSION

In Setting 1, we compare VisionPro Deep Learning results with other open-source Deep Learning architectures and see that it beats those other networks in the overall F-scores.

In Setting 2, approximately 9,000 images are reduced from the training set. We see that the performance of all the models only reduces slightly in this setting.

In Setting 3, approximately 35,000 images are reduced from the original COVIDx-CT [23][29] dataset. Even though the performance of all open-source models decreases on such significant reduction of training images, VisionPro Deep Learning still produces similar results as in Setting 1 or 2, with overall F-scores still in 99%. This highlights the stability and robustness of the VisionPro Deep Learning GUI, even though the training images are reduced significantly. This robust performance against dramatic decreases in the training data (while increasing the test data at the same time) proofs that the model also should perform best in the field.

Among the open-source models, the Xception [28] architecture almost always performs better than ResNet [19], DenseNet [20] and Inception [27].

The total time needed to do the study on Python when compared with VisionPro Deep Learning varies a lot. In python, creating and debugging the code from scratch, that is, to load, preprocess, and finally train the deep learning network to get the best results can take of timeframe ranging from somewhere around 15 to 20 days, or even more depending on the proficiency of the programmer. In VisionPro, the images just need to be loaded and labeled, and with the click of a button, the training can be started. The entire process of loading the images, labelling the images, training the network and getting the final results takes a timeframe of 2 to 3 days. This reduction in time, can prove to be very beneficial in a clinical environment.

## CONCLUSION

In this study we use COGNEX's Deep Learning Software- VisionPro Deep Learning (version 1.0) and compare its performance to other state of the art Deep Learning architectures. VisionPro Deep Learning has an intuitive GUI making the software very easy to use. Building applications requires no coding skills in any programming language. Little to no preprocessing of the images is required, which also decreases the development time. Imbalanced data is automatically balanced within the software. Once the images are loaded into VisionPro Deep Learning and the right tool is selected, the Deep Learning training can start. After completion of training, it outputs a confusion matrix, along with the various important metrics, such as precision, recall and F-score. Additionally, a report can be generated that identifies all misclassified images. This makes it particularly suitable for radiologists, hospitals, and research workers to harness the power of Deep Learning without advanced coding knowledge.

Moreover, as the results from this study indicates, the Deep Learning algorithms in VisionPro Deep Learning are robust and comparable or even better than the various state of the art algorithms available today. Also, even with significantly fewer images in the training set, even though the performance of all the open-source models fall, VisionPro Deep Learning still gets F-scores of 99%. A heatmap can be generated to showcase exactly where the model is focusing on while making the predictions. On all the three settings, VisionPro Deep Learning achieves the highest overall F-scores, surpassing the results of the various open source architectures.

This software is by no means a stand-alone solution in the identification of images COVID-19 from CT scans , but can aid radiologists and clinicians to achieve a faster and understandable diagnosis using the full potential of Deep Learning, without the prerequisite of having to code in any programming language.

**ACKNOWLEDGEMENT**

We will like to thank COGNEX for providing their latest Deep Learning software for testing, and University of Waterloo, along with Darwin AI for collecting the CT images and creating the dataset and for providing the python scripts for generating the COVIDx-CT [23] dataset.

**AUTHOR'S CONTRIBUTIONS**

Arjun Sarkar wrote the manuscript. Arjun Sarkar, Joerg Vandenhirtz, Jozsef Nagy and David Bacsa conducted the experiment. Arjun Sarkar, Joerg Vandenhirtz, Jozsef Nagy, David Bacsa, Mitchell Riley analyzed the results. All authors reviewed the manuscript.

**ADDITIONAL INFORMATION**

Arjun Sarkar, Joerg Vandenhirtz, Jozsef Nagy, David Bacsa, Mitchell Riley are affiliated with COGNEX and COGNEX funded the research in this paper.